\documentclass[fleqn,usenatbib]{mnras}

\usepackage{newtxtext,newtxmath}

\usepackage[T1]{fontenc}
\usepackage{ae,aecompl}

\usepackage{graphicx}	
\usepackage{amsmath}	
\usepackage{amssymb}	
\usepackage{float}
\usepackage{pdflscape}
\usepackage{soul}
\usepackage[usenames,dvipsnames]{color}
\usepackage[normalem]{ulem}

\newcommand{\Msun}{\mathrm{M}_\odot}

\newcommand{\pkg}[1]{\texttt{#1}}
\newcommand{\single}[1]{\overline{#1}}
\newcommand{\double}[1]{\widehat{\overline{#1}}}
\newcommand{\model}[1]{\texttt{#1}}

\newcommand{\smag}{$C_\mathrm{s}$ }
\newcommand{\smagend}{$C_\mathrm{s}$}
\newcommand{\xmark}{$\times$}
\newcommand{\mvec}[1]{\boldsymbol{#1}}

\title[Dynamic Turbulent Diffusion]{Dynamic Localised Turbulent Diffusion and its Impact on the Galactic Ecosystem}

\author[Rennehan et al.]{
Douglas Rennehan,$^{1}$\thanks{E-mail: rennehan@uvic.ca (DR)}
Arif Babul,$^{1}$
Philip F. Hopkins,$^{2}$
Romeel Dav\'{e}$^{3,4,5}$
\newauthor \ and Belaid Moa$^{6}$
\\
$^{1}$Department of Physics \& Astronomy, University of Victoria, BC V8X 4M6, Canada\\
$^{2}$TAPIR, Mailcode 350-17, California Institute of Technology, Pasadena, CA 91125, USA\\
$^{3}$Department of Physics and Astronomy, University of the Western Cape, Bellville, Cape Town 7535, South Africa\\
$^{4}$South African Astronomical Observatories, Observatory, Cape Town 7925, South Africa\\
$^{5}$Institute for Astronomy, Royal Observatory, Edinburgh EH9 3HJ, UK\\
$^{6}$Compute Canada/WestGrid/University Systems, University of Victoria, BC V8P 5C2, Canada
}

\date{Accepted 2018 December 07. Received 2018 November 27; in original form 2018 July 30}

\pubyear{2018}

\begin{document}
\label{firstpage}
\pagerange{\pageref{firstpage}--\pageref{lastpage}}
\maketitle

\begin{abstract}
Modelling the turbulent diffusion of thermal energy, momentum, and metals is required in all galaxy evolution simulations due to the ubiquity of turbulence in galactic environments.  The most commonly employed diffusion model, the Smagorinsky model, is known to be over-diffusive due to its strong dependence on the fluid velocity shear.  We present a method for dynamically calculating a more accurate, locally appropriate, turbulent diffusivity:  the dynamic localised Smagorinsky model.  We investigate a set of standard astrophysically-relevant hydrodynamical tests, and demonstrate that the dynamic model curbs over-diffusion in non-turbulent shear flows and improves the density contrast in our driven turbulence experiments.  In galactic discs, we find that the dynamic model maintains the stability of the disc by preventing excessive angular momentum transport, and increases the metal-mixing timescale in the interstellar medium. In both our isolated Milky Way-like galaxies and cosmological simulations, we find that the interstellar and circumgalactic media are particularly sensitive to the treatment of turbulent diffusion.   We also examined the global gas enrichment fractions in our cosmological simulations, to gauge the potential effect on the formation sites and population statistics of Population III stars and supermassive black holes, since they are theorised to be sensitive to the metallicity of the gas out of which they form.  The dynamic model is, however, not for galaxy evolution studies only.  It can be applied to all astrophysical hydrodynamics simulations, including those modelling stellar interiors, planetary formation, and star formation. 
\end{abstract}




\section{Introduction}
\label{sec:introduction}

Galaxies form at the confluence of gas streams and cooling flows at the centres of virialized halos and evolve via a constant exchange of baryons with their environments.  Despite significant recent progress on understanding the details of this picture, developing predictive models for the evolution and observed properties of galaxies has proven to be an immense challenge (see, for example, \citealt{Guedes2011, Hopkins2014, Vogelsberger2014, Schaye2014, Genel2016, Dave2017}; -- see also \citealt{Somerville2015} and \citealt{Naab2016} for a recent review and additional references).  The problem lies in the large number of complex interconnected processes involved, and the huge dynamic range in spatio-temporal scales over which they operate.  

One important interplay involves the interstellar medium (ISM), the gas that permeates a galaxy and provides fuel for star formation, and circumgalactic medium (CGM), the gas that cocoons the galaxy.  The amount of gas in the CGM and the efficiency with which it can cool, fall into the galaxy and replenish the ISM, are important variables in setting the duration and the rate of star formation \citep{Somerville2015}.  Stellar winds and supernova explosions (SNe) -- processes directly related to star formation -- provide competition for gas cooling \citep{Springel2003, Oppenheimer2008}.   These deposit energy and momentum into the ISM, engendering outflows of gas.   If these outflows are sufficiently powerful, they not only heat the CGM, but also expel most of the CGM from the galaxy's halo.   This ongoing competition between gas into and out of galaxies provides a basic framework for understanding a number of observed properties \citep{Shen2012, Crain2013, Christensen2016, Oppenheimer2016, Sokolowska2016}.

The cooling efficiency and ionisation state of the gas in the CGM depend sensitively not only on the spatial injection and redistribution of thermal energy and momentum \citep{Suresh2017}, but also on the metals that are transported from the galaxy via the galactic outflows \citep{Dave2006, Oppenheimer2006, Finlator2008, Hani2018}.  Metals, although negligible in terms of mass fraction, play an out-sized role in galaxy evolution because they can dramatically alter the CGM's cooling profile \citep{Van2012, Oppenheimer2013, Sokolowska2017} and, hence, the delicate balance between gas in- and out- flow.  

In addition to cooling, the redistribution of metals can impact other processes, such as the sites and formation history of the putative Population III (Pop III) stars and supermassive black holes (SMBHs).   Pop III stars are associated with star formation involving near-pristine gas with an upper limit on metallicities somewhere in the range $[Z] \sim -6$ to $[Z] \sim -3$ \citep{Sarmento2016}.  As for SMBHs, a number of authors (e.g. \citealt{Volonteri2010} and for use in recent simulations \citealt{Tremmel2017}) postulate that they form via direct collapse of gas clouds with metallicity $[Z] \lesssim -4.0$, with the idea that very low metallicity would prevent the gas from cooling rapidly and fragmenting into Pop III stars during collapse.  It is therefore crucial to identify which physical processes redistribute thermal energy, momentum, and metals in galactic environments spatially, and to include these accurately in the numerical galaxy evolution experiments.

One critical, often overlooked, redistribution mechanism is gas turbulence. Turbulence occurs when inertial forces dominate viscous forces in a gaseous environment, and kinetic energy injected on large scales cannot immediately dissipate as heat.  This leads to the formation of a kinetic energy cascade, as coherent turbulent eddies on large scales spawn eddies on successively smaller scales, until the energy thermalises.  Galactic environments, for example, are expected to be highly turbulent \citep{Evoli2011, Iapichino2013}.  In the case of the CGM, this is strongly suggested by the kinematic complexity revealed by absorption and emission line measurements \citep{Tumlinson2017} and it has long been recognized that the cold ISM is also highly turbulent.  The susceptibility of a medium to turbulence is quantified by its dimensionless Reynolds number\footnote{Defined as the ratio of inertial forces to viscous (dissipative) forces.}, $Re$.  The Reynolds number for the cold ISM has been estimated to be as high as $Re \sim 10^7$ \citep{Elmegreen2004}, whereas the onset of turbulence usually occurs at $Re \sim 10^3$.  The Reynolds number is also a measure of the separation of scales in the energy cascade and, in the case of incompressible turbulence,  $L / \eta \sim Re^{3/4}$, where $L$ is the kinetic energy injection scale, and $\eta$ is the dissipation scale\footnote{In compressible turbulence, the scaling is much steeper (see \citealt{Kritsuk2007, Federrath2013a}).}.  Therefore the degrees of freedom in a 3D simulation scales as $Re^{9/4}$, and simulating a $Re \sim 10^{7}$ flow would require $10^{15}$ fluid elements!  Contemporary cosmological simulations have reached $\sim 10^{12}$ fluid elements and a dynamic range typically of the order $\sim 10^{6}$ \citep{Somerville2015}, with the smallest scale being the resolution limit $h$.  Therefore all cosmological simulations that involve turbulence -- independent of hydrodynamical method -- have a natural cut-off scale $h$, in the range $\eta \ll h \ll L$, where discretisation truncates the turbulent cascade.

Physically, small-scale turbulent fluctuations, by promoting mixing, provide a transport mechanism for the fluid properties such as momentum, thermal energy, and metals.  In numerical simulations, this implies that turbulence on scales smaller than $h$ can potentially impact the resolved properties of the flow and consequently must be accounted for \citep{Germano1991}.  The crux of the issue is that, as stated above, the kinetic energy flux down the turbulent cascade is truncated at $h$ in the simulations whereas, in reality, the kinetic energy cascade should continue to smaller scales until it is dissipated.  Numerical simulations break the physical coupling between the scales that are resolved ($> h$) and unresolved ($< h$), and, therefore, they require models of (i) the kinetic energy flux from the resolved to the unresolved scales, (ii) the effect of unresolved eddies on the resolved scales, and (iii) the transport properties of kinetic energy on unresolved scales.  Typically, this is done using the \textit{turbulent eddy-diffusion models} that treat sub-grid turbulent eddy motion as a diffusive process.  Several implementations of this approach have been proposed (cf. \citealp{Schmidt2011, Wadsley2017, DiMascio2017}; -- see  \citealp{Sagaut2006} and \citealp{Garnier2009} for extensive lists);  nonetheless, many cosmological hydrodynamical simulation studies assume that numerical diffusion adequately accounts for sub-grid turbulent transport \citep{Schmidt2015}.  Numerical diffusion, however, can lead to diffusive behaviour that poorly represents turbulent flow statistics \citep{Sagaut2006}.
 
Extensive effort has been devoted toward developing models for treating turbulent diffusion in Eulerian cosmological simulations, which employ grids to discretise the fluid.  Here we refer the reader to \citet{Scannapieco2008, Pan2013, Federrath2013a, Schmidt2013, Schmidt2015, Semenov2016, Sarmento2016}.  In this paper we focus on the Lagrangian hydrodynamical approach.

Within the Lagrangian framework, the fluid equations of motion are approximated by tracking individual fluid elements as they move with the flow.  Commonly used Lagrangian methods in computational cosmology include smoothed particle hydrodynamics (SPH) \citep{Lucy1977, Gingold1977, Hernquist1989}, moving-meshes (MM) \citep{Springel2010}, and higher-order mesh-free methods (MF) \citep{Lanson2008a, Lanson2008b, Gaburov2011, Hopkins2015a} -- see \cite{Springel2010} and \cite{Hopkins2015a} for extensive discussion.  

SPH has no inherent diffusion and therefore, by construction, explicitly requires additional transport terms, one benefit of which is full control over the strength of mixing.  In MM and MF methods, numerical diffusion arises as a by-product of the numerical scheme used to solve the Riemann problem between fluid elements. However, as noted above, inherent numerical diffusion does not, in general, reproduce the correct turbulent properties of a simulated flow.  The key problem here is: how can we address this? Or, more precisely, how do we model the interaction between the unresolved and the resolved scales?

A common approach is to assume that the interaction between the resolved and the unresolved scales reduces to a local transfer of kinetic energy from the large to the small scales.  If we additionally assume that the kinetic energy transfer mechanism is analogous to a diffusive process, where kinetic energy is redistributed on progressively smaller scales via momentum diffusion, then we can treat the action of unresolved eddies in a similar fashion to molecular viscosity.  Under this equivalency, on scales close to the resolution scale $h$, the unresolved eddies extract kinetic energy from the resolved flow via momentum diffusion, and simultaneously allow for the dissipation of kinetic energy.  This is the \textit{eddy-viscosity} hypothesis, and is fundamentally a model assumption that describes the simplest view of the interactions between the scales.  The action of the unresolved eddies can then be modelled by a viscous term in the fluid equations of motion -- diffusing momentum and dissipating kinetic energy \citep{Pope2000}.  The above assumptions lead to a simple model but it does not provide guidance about the appropriate choice for the effective diffusivity/viscosity.

In a turbulent cascade, the diffusive action of the eddies depends on their velocity and length scale.  On the sub-grid level, this involves estimating the velocity scale that transports fluid properties over the resolution scale $h$ \citep{Wadsley2008, Greif2009a}.  In the simplest model, the Smagorinsky model \citep{SMAGORINSKY1963}, this velocity is assumed to be proportional to the gradients of the velocity field.   In this model, the sub-grid eddies both diffuse (characterised by diffusivity $D$) and dissipate via eddy viscosity, $\nu_\mathrm{sgs}$, given by $\nu_\mathrm{sgs} = D = ($\smagend$ h)^2 \left | S^* \right |$, where $|S^*|$ is the norm of the trace-free shear tensor, and \smag is the Smagorinsky model constant.  Contemporary fluid mechanics literature notes that in general \smag needs to be tuned to a value between $0.1$ and $0.2$ for optimal results under different flow conditions \citep{Garnier2009}.  

The main advantage of the Smagorinsky model is its simplicity and as a result, many researchers have started to incorporate this model into their cosmological simulation codes -- specifically as a model for the diffusivity when treating thermal energy and metal mixing \citep{Shen2010, Brook2014, Williamson2016a, Tremmel2017, Sokolowska2017, Escala2017}.  The model, however, has some drawbacks:  (i) A single valued $C_s$ is incapable of correctly describing different types of turbulent flows.  Studies show that the Smagorinsky model introduces too much diffusion into the flow in almost all cases except for homogeneous, isotropic turbulence \citep{Garnier2009}.  And, (ii) the sub-grid eddy viscosity does not vanish for laminar shear flows where there ought to be no diffusion due to turbulence.  Over-diffusion is especially worrisome given the  push to resolve the multiphase structure in the ISM and CGM at greater levels, and the recent results that differential, localised metal mixing could change our understanding of galactic chemical evolution \citep{Emerick2018}.

As noted above, most implementations of the Smagorinsky model only consider thermal energy and metal diffusion, but not momentum diffusion because the latter drawback above is a concern for differentially rotating structures, such as galactic discs.  Specifically, it results in undesired angular momentum transport and, consequently, unphysical flows in the discs.  Most galaxy evolution studies ignore momentum diffusion in order to avoid this viscous instability.  The crux of the problem is that a constant \smag cannot automatically adjust to either the local conditions or the changing character of the flow with time.  This led \cite{Germano1991} to propose the dynamic Smagorinsky model where \smag is a function of space and time, i.e. \smag$=$ \smagend$(\mvec{x}, t)$.  

The dynamic Smagorinsky model has, to our knowledge, not yet been implemented and investigated in cosmological simulations.  The model has, however, been validated extensively in the fluid simulation community by comparing to the results of standard numerical tests and experiment data, and has been shown to improve upon the constant-coefficient model \citep{Kleissl2006, Kirkpatrick2006, Benhamadouche2017, Lee2017, Kara2018, Taghinia2018}.  In the dynamic model, the sub-grid properties of a turbulent fluid are computed under two assumptions\footnote{The two assumptions are often combined together and referred to as the scale-similarity hypothesis.}: (i) the behaviour of the largest \textit{unresolved} eddies is entirely determined by their interactions with the eddies on the smallest \textit{resolved} scales, and (ii) these interactions are analogous to those between the fluid motions on the smallest resolved scale and the motions on larger scales.  In practice, determining the characteristics of these interactions involves filtering (or smoothing) the resolved velocity field on two different scales.  When the sub-grid turbulent properties are calculated based on the local fluid properties, \smagend$(\mvec{x}, t)$ reduces to zero (i.e. the eddy viscosity/diffusivity vanishes) in non-turbulent (or laminar) shear flows \citep{Piomelli1994}. Consequently, this allows for the self-consistent treatment of momentum diffusion, along with thermal energy and metal diffusion, in numerical studies of cosmic baryons and galaxy evolution.

In this study, we introduce an implementation of the dynamic Smagorinsky model for the first time in Lagrangian astrophysical simulations, focusing on the higher order MF approach.  We discuss the implementation details in Section \ref{sec:methods}.  In order to test the impact of dynamic localised turbulent mixing on the galactic ecosystem we run a series of hydrodynamical and physical experiments relevant to galaxy evolution.  In Section \ref{sec:hydrotests} we show the results for a set of standard hydrodynamic tests and explicitly check on the extent of sub-grid diffusion in laminar shear flows.  We then go on to examine the effects of dynamic mixing in an isolated disc galaxy in Section \ref{sec:isogalaxy}, followed by a set of cosmological simulations in Section \ref{sec:cosmo}.  We investigated SPH and performed all of the experiments presented below, but we do not include them in this paper because the results are qualitatively similar to the MF results, as we mention in Section \ref{sec:conclusions}.

\section{Methods}
\label{sec:methods}

\subsection{Hydrodynamics}
\label{sec:gizmo}

In order to test the impact and robustness of the localised turbulent diffusion model, we employ a modified version of the \pkg{GIZMO} gravity plus hydrodynamics solver code \citep{Hopkins2015a}.  \pkg{GIZMO} builds on the \pkg{GADGET-3} code base \citep{Springel2005}, with improvements in numerical accuracy and includes an implementation of the novel mesh-free finite mass (MFM) method, in addition to various implementations of smoothed particle hydrodynamics (SPH) methods.  

The MFM method evolves the fluid equations of motion in a Lagrangian manner similar to SPH. However, while the fluid mass elements in SPH are discretised into particles and their motions are determined by fluid properties smoothed over neighbouring particles, the conservation laws in MFM are evolved by calculating the flux of basic variables between neighbouring particles\footnote{We refer to any fluid element as a particle, for simplicity. Fluid elements in the MFM method are not particles in an SPH sense, and are defined by the effective geometrical faces moving along lines connecting cells enclosing a finite mass.}. These fluxes depend on the effective face area between the two particles and are determined by solving the Riemann problem along the line connecting them.  This removes the need for additional terms, such as artificial viscosity and conductivity as is necessary in SPH to ensure proper treatment of shocks.  The numerical Riemann solvers have inherent numerical dissipation\footnote{Riemann solver dissipation arises from the high, even-order, truncated terms in the Taylor series expansion of the basic variables in the conservation equations.}, hence improved shock capturing capabilities. We, therefore, focus on the use of the MFM method in our investigation of the dynamic diffusion model. For a thorough exposition of \pkg{GIZMO} and an extensive comparison between MFM and SPH, see \cite{Hopkins2015a}.

\subsection{Sub-grid turbulent diffusion terms}
\label{sec:additionalterms}

\pkg{GIZMO} solves the conservation equations for momentum, energy, and mass using the MFM method, and like other hydrodynamic methods, it is limited in resolution down to a scale $h$. The minimum resolution limits the ability to resolve high Reynolds number flows down to the viscous dissipation scale, impacting its ability to resolve the turbulent cascade.  As mentioned in Section \ref{sec:introduction}, the interaction of the resolved and unresolved scales must be modelled.  These models apply to the additional sub-grid scale terms that appear in the equations of motion when treating discretisation as a filtering process \citep{Sagaut2006, Garnier2009, Schmidt2015}.  In this section, we detail the origin of the sub-grid scale terms and discuss which terms we include in the \pkg{GIZMO} code.

Discretisation of the conservation equations is equivalent to applying a low-pass filter, damping out high frequency turbulent fluctuations.  When we discuss filtering, we refer to the definition of a general filtered scalar field $f(\mvec{x})$,

\begin{equation}
\single{f}(\mvec{x}) \equiv \int_D f(\mvec{x}')G(|\mvec{x}' - \mvec{x}|, \single{h}) \mvec{dx}',
\label{eq:filterdef}
\end{equation}

\noindent where $G(|\mvec{x}'-\mvec{x}|, \single{h})$ is a filter function.  $\single{h}$ is the characteristic size of the filtering operation below which fluctuations are damped (essentially the resolution scale, in the present context), $\mvec{dx}'$ is a volume element, and the integral is evaluated over the entire domain.  We discuss our filtering implementation in more detail, in Section \ref{sec:dynamicmodel}. We now apply this equation to the conservation equations in order to see that additional terms in the hydrodynamical equations are required.

The momentum conservation equation for a compressible fluid follows\footnote{We follow this Einstein notation throughout this paper.} \citep{Landau1987}, 

\begin{equation}
\frac{\partial}{\partial t}(\rho u_i) + \frac{\partial}{\partial x_j}[\rho u_i u_j + p\delta_{ij}] = 0,
\label{eq:navierstokes}
\end{equation}

\noindent where $u_i = u_i(\mvec{x},t)$ is the fluid velocity vector in the $i = \{x,y,z\}$ direction, $\rho = \rho(\mvec{x}, t)$ is the fluid density, and $p = p(\mvec{x}, t)$ is the pressure.  When we filter the momentum equation, assuming the filtering operation and derivatives commute, we end up with an extra term $\tau_{ij}$,

\begin{equation}
\frac{\partial}{\partial t}(\overline{\rho} \tilde{u_i}) + \frac{\partial}{\partial x_j}[\tau_{ij} + \single{p}\delta_{ij} + \single{\rho}\tilde{u}_i\tilde{u}_j] = 0,
\label{eq:addtermnavierstokes}
\end{equation}

\noindent where we have also switched to density weighted variables such that $\tilde{u}_i = \single{\rho u_i} / \single{\rho}$. $\tau_{ij}$ is the \textit{subgrid-scale turbulent stress tensor} or \textit{residual stress tensor} and is defined as,

\begin{equation}
\tau_{ij} \equiv \single{\rho} (\widetilde{u_i u_j} -  \tilde{u}_i\tilde{u}_j).
\label{eq:sgsstresstensor}
\end{equation}

\noindent This term must be modelled because $\widetilde{u_i u_j}$ is unknown at the time of simulation, i.e., the system of equations is not closed.  A common model, or closure, involves the eddy-viscosity assumption where the sub-grid scales impart a momentum flux on the resolved scales that is linearly dependent on the rate of strain of the resolved scale,

\begin{equation}
\tau_{ij} = -2 \single{\rho} \nu_\mathrm{sgs} \widetilde{S^*}_{ij},
\label{eq:eddyviscosity}
\end{equation}

\noindent where $\nu_\mathrm{sgs}$ is the sub-grid eddy viscosity, and $\widetilde{S^*}_{ij}$ is the trace-free resolved rate of strain tensor.  The same logic can be applied to any of the conservation equations and any filtered multiplicative terms require modelling.  In the derivations below, we model (as in equation~\ref{eq:eddyviscosity} above) the unknown terms under the assumption that they act as diffusive processes.

There is a similar term when filtering the total energy equation ($e$ is the specific total energy),

\begin{equation}
\frac{\partial}{\partial t}(\rho e) + \frac{\partial}{\partial x_j}[\rho u_j e + u_j p] = 0.
\end{equation}

\noindent leads to additional terms,

\begin{equation}
\frac{\partial}{\partial t}(\overline{\rho} \tilde{e}) + \frac{\partial}{\partial x_j}[Q_j + P_j + \overline{\rho} \tilde{u}_j \tilde{e} + \overline{u}_j \overline{p}] = 0,
\end{equation}

\noindent where $e = \theta + \frac{1}{2} |\mvec{u}|^2$ ($\theta$ is the specific internal energy).  We also have defined:

\begin{equation}
Q_j \equiv \overline{\rho} (\widetilde{u_j e} - \tilde{u}_j \tilde{e}), \\
P_j \equiv \overline{u_j p} - \overline{u}_j \overline{p}.
\end{equation}

\noindent In this study, we ignore the term associated with pressure, $P_j$, and the term in $Q_j$ that arises from $\frac{1}{2}|\mvec{u}|^2$ in $e$, and focus on the application of the dynamic model to the sub-grid momentum term working in concert with the widely employed thermal energy term \citep{Shen2013, Brook2014, Tremmel2017, Wadsley2017},

\begin{equation}
q_j \equiv \overline{\rho} (\widetilde{u_j \theta} - \tilde{u}_j \tilde{\theta}) = -\single{\rho} \nu_\mathrm{sgs} \frac{\partial \tilde{\theta}}{\partial x_j}.
\label{eq:heatflux}
\end{equation}

Not only are there additional terms for momentum and thermal energy, but any scalar quantities, such as the concentration of different metal species, in the gas should be transported in a turbulent flow. In order to model this, we treat metal concentrations, $\phi_z$, as passive scalars that obey a diffusion equation \citep{Pope2000, Shen2010},

\begin{equation}
\frac{\partial \phi_z}{\partial t} = \frac{\partial}{\partial x_j} \bigg( \overline{\rho} \nu_\mathrm{sgs} \frac{\partial \phi_z}{\partial x_j} \bigg).
\label{eq:genericdiff}
\end{equation}

\noindent For a detailed description of the incorporation of these fluxes into \pkg{GIZMO}, see \cite{Hopkins2016b}.

Throughout this paper, we refer to the action of the terms in equations~(\ref{eq:eddyviscosity}), ~(\ref{eq:heatflux}), and ~(\ref{eq:genericdiff}) as \textit{turbulent diffusion} because they contribute to the conservation equations as $\nabla^2 f$, where $f$ is the flux quantity.  Additionally, when we mention energy diffusion, we are referring to the term $\mvec{\nabla}\cdot\mvec{q}$ (thermal energy diffusion), and, similarly, when we mention momentum diffusion we are referring to the action of the stress tensor through $\mvec{\nabla}\cdot\mvec{\tau}$, along with the corresponding kinetic energy dissipation.

\subsection{Diffusivity}
\label{sec:smagimplementation}

Physically, turbulent mixing can be modelled as a diffusive process with diffusivity $D$ and, in the simplest model, the fluid properties are assumed to mix over the resolution scale $h$ with a velocity $h|S^*|$, where $|S^*|$ is the norm of the trace-free shear tensor.  This is, as we mentioned in the introduction, the Smagorinsky model and the corresponding diffusivity is parametrised as $D = \nu_\mathrm{sgs} =(C_\mathrm{s} h)^2 |S^*|$.

The Smagorinsky model inherently assumes that the kinetic energy transfer rate down the turbulent cascade is equal on all scales, and is equal to the physical dissipation rate (i.e. the flow is in local equilibrium).  In simulations, the resolution scale $h$ inhibits kinetic energy from moving to progressively smaller scales, and results in a build-up of kinetic energy at the resolution scale -- so long as numerical dissipation cannot extract kinetic energy sufficiently rapidly.  The Smagorinsky model combined with the local equilibrium assumption only provides a model for the turbulent stress, $\tau_{ij}$, and dissipation in the flow, $\Sigma = \tau_{ij} \tilde{S}^*_{ij}$, and ignores the additional terms we discuss in Section~\ref{sec:additionalterms}.  In order to consistently model all of the energetic terms, one must relax the local equilibrium assumption and follow the sub-grid kinetic energy, $K$, directly.  It is possible to derive a one-equation model for $K$ that includes a third-order term for the transport of $K$ on sub-grid scales \citep{Schmidt2015}, and self-consistently follows all of the sub-grid quantities.  Each sub-grid term can then be modelled using diffusive terms similar to the models in Section~\ref{sec:additionalterms}, in order to close the system of equations.

If the Smagorinsky model only considers the turbulent stress, is it then valid to apply this model (as we have done following \citealt{Shen2010}) to the thermal energy and metal sub-grid terms?  In order for this to be possible, the local equilibrium condition must be approximately true in the regime of interest.  We are specifically interested in cosmological-scale gas, and \cite{Schmidt2016} show that the local equilibrium condition holds -- on average -- in a cosmological-scale volume.  Introducing the dynamic model on top of these approximations further supports our model assumption, because the dynamic model inherently accounts for the deviations from local equilibrium.

The Smagorinsky model diffusivity is parametrised in \pkg{GIZMO} for a particle $a$ as,

\begin{equation}
\label{eq:smaggizmo}
D_{a} = \rho_{a} (C_\mathrm{s}  h_{a})^2 |S^*|_{a},
\end{equation}

\noindent where \smag is the Smagorinsky constant, $ h_{a} $ is the mean inter-particle spacing in the kernel, and $ |S^*|_{\mathrm{a}} $ is the magnitude of the trace-free symmetric shear tensor.  Note that we absorb the densities $\overline{\rho}$ from Section~\ref{sec:additionalterms} into $\nu_\mathrm{sgs}$ via $\rho_a$.  $D_\mathrm{a}$ is used in the diffusion equations for thermal energy, momentum, and metal mass fractions as described in \cite{Hopkins2016b}.  There are a myriad of values quoted for \smag in the literature (see Section 5 of \cite{Sagaut2006} for an extensive list), but we choose the value calibrated for fully-developed isotropic turbulence, \smag$= 0.2$ \citep{Clark1979}, because the Smagorinsky model was developed for this specific regime.  

In order to compare our \smag with other values in the literature, it is important to consider the definitions of the quantities in equation~(\ref{eq:smaggizmo}).  In many SPH studies, the length scale in the diffusion coefficient definition is taken as the kernel-support radius, $h_\mathrm{SPH}$, which is the maximum extent from a particle that gives a non-zero weight.  In contrast, as we mentioned above, we employ the mean inter-particle spacing within the kernel or $h_a \approx 0.5 h_\mathrm{SPH}$.  Additionally, some studies use $C \equiv \sqrt{2} $\smagend$^2$ whereas for this study $\sqrt{2}$ is absorbed into our definition of the norm of the shear tensor (see equation~\ref{eq:magsheargizmo} below), closely following the fluid simulation literature \citep{Piomelli1994}.  Using these definitions, our adopted value of \smag is lower than those quoted in the astrophysics literature.  For example, the value $C = 0.05$ (see \citealt{Shen2010, Shen2013, Brook2014}), corresponds to \smag$=0.37$ whereas $C = 0.03$ corresponds to \smag$=0.29$ (see \citealt{Wadsley2017}).  However, our \smag is higher than the value recently calibrating from studying metal mixing in dwarf galaxies, where \cite{Escala2017} found \smag$=0.046$ reproduced more realistic stellar metal distribution functions via supersonic mixing in the interstellar medium. 

We compute the trace-free symmetric shear tensor via the high-order accurate gradient estimators in \pkg{GIZMO},
 
\begin{equation}
\label{eq:symmetricsheargizmo}
S_{ij}^* = \frac{1}{2}\bigg(\frac{\partial u_i}{\partial x_j} + \frac{\partial u_j}{\partial x_i}\bigg) - \frac{1}{3}\delta_{ij}\frac{\partial u_k}{\partial x_k},
\end{equation}

\noindent where $ u_i $ is the fluid velocity vector and $x_i$ the spatial coordinate, and $ i, j = \{x, y, z\} $.  The magnitude of equation~(\ref{eq:symmetricsheargizmo}) is implemented using the Frobenius norm \citep{Piomelli1994},

\begin{equation}
\label{eq:magsheargizmo}
|S^*| = \sqrt{2 S_{ij}^* S_{ij}^*}.
\end{equation}

We note that in \pkg{GIZMO}, \cite{Escala2017} impose an ad hoc cap on the diffusivity based on the expected maximum mass flux between resolution elements.  The cap does not, however, mitigate the fact that the constant Smagorinsky model induces diffusion whenever there is shear, regardless of whether the fluid is laminar or turbulent.  It is only intended to prevent unphysical mass/energy transport that can potentially arise due to the excessive diffusivity of the Smagorinsky model.  We adopt the same limiter in this study, but note that the diffusivity rarely reaches the maximum limit.

\subsection{Dynamic Model}
\label{sec:dynamicmodel}

The Smagorinsky model provides an approximate model of sub-grid mixing for fully developed, homogeneous turbulence but it is far too simple for complex flows. In fact, in laminar shear flows, the constant-coefficient Smagorinsky model predicts a non-zero diffusivity through its dependence on the shear strength (equation~\ref{eq:smaggizmo}).  However, in this situation, the diffusivity should be zero since the fluid is not turbulent.  In more complex flows, such as those in astrophysical contexts, the value of the constant ought to depend on the spatio-temporal coordinates $ C_s = C_s(\mvec{x}, t) $ \citep{Germano1991}. \cite{Piomelli1994} showed that by assuming scale-similarity (cf. Section \ref{sec:introduction}), the local Smagorinsky constant in a neighbourhood can be calculated at each point (at a fixed simulation timestep) as follows:

\begin{equation}
\label{eq:dynamicc}
C_{\mathrm{dyn}}(\mvec{x}) = C_s^2 = -\dfrac{1}{2} \dfrac{(L_{ij} - 2 \widehat{C_{\mathrm{dyn}}^{p}\beta_{ij}})\alpha_{ij}}{\alpha_{mn}\alpha_{mn}}.
\end{equation}

\noindent Here $ L_{ij} $ is the Leonard tensor, 

\begin{equation}
\label{eq:leonardtenstor}
L_{ij} = \widehat{\single{u}_i \single{u}_j} - \double{u}_i \double{u}_j,
\end{equation}

\noindent $C_{\mathrm{dyn}}^{p}$ is the value of $C_{\mathrm{dyn}}$ at the previous timestep, and $ \alpha_{ij} $ and $ \beta_{ij} $ are defined as,

\begin{equation}
\label{eq:alphaij}
\begin{split}
\alpha_{ij} \equiv \widehat{h}^2 |\double{S^*}| \double{S^*}_{ij},\\
\beta_{ij} \equiv \single{h}^2 |\single{S^*}| \single{S^*}_{ij}.
\end{split}
\end{equation}

\noindent Here $\single{f}$ represents a filtering (or smoothing) operation on $f$ over a length-scale $\single{h}$, and $\widehat{f}$ represents smoothing on a scale $\widehat{h}$.  $\single{h}$ usually is equated with the lowest resolvable scale, and extensive work has been done in the fluid simulation community to show that the optimal value for $\widehat{h}$ is $\widehat{h} = 2 \single{h}$ \citep{Germano1991, Piomelli1994, Spyropoulos1996, Schmidt2006, Grete2017}.  For this work, we choose $\single{h}$ to be the compact support radius of the kernel, and $\widehat{h} = 2 \single{h}$.  

More precisely, we smooth a scalar field $ f(\mvec{x}) $ by convolving it with the filter function $ G(|\mvec{x}' - \mvec{x}|, \single{h}) $ over the domain\footnote{Our filtering implementation naturally density-weights quantities because we follow the hydrodynamical weighting scheme.},

\begin{equation}
\label{eq:convolution}
\single{f}(\mvec{x}) = \int_D f(\mvec{x}')G(|\mvec{x}' - \mvec{x}|, \single{h}) \mvec{dx}'.
\end{equation}

\noindent This is similar to the SPH method of interpolating a scalar function, with $G(|\mvec{x}' - \mvec{x}|, \single{h})$ sharing the same properties as the smoothing kernel $W(|\mvec{x}' - \mvec{x}|, \single{h})$,

\begin{equation}
\label{eq:filterproperties}
\begin{split}
\int G(|\mvec{x}' - \mvec{x}|, \single{h})\mvec{dx}' = 1,\\
\lim_{\single{h}\to 0} G(|\mvec{x}' - \mvec{x}|, \overline{h}) = \delta(|\mvec{x}' - \mvec{x}|).
\end{split}
\end{equation}

\noindent The MFM method employs a similar technique for evaluating integrals and in order to be consistent, we choose $G = W$. The integral in equation~(\ref{eq:convolution}) is expensive, but we simplify the computation using XSPH smoothing \citep{Monaghan1989, Monaghan2005, Monaghan2011},

\begin{equation}
\label{eq:xsphsmoothing}
\overline{f}(\mvec{x}) = f(\mvec{x}) + \epsilon \int_D (f(\mvec{x}') - f(\mvec{x}))W(|\mvec{x}' - \mvec{x}|, \overline{h})\mvec{dx}'.
\end{equation}

\noindent  Following \cite{Monaghan2011}, the Fourier coefficients $a_\mathrm{k}$ of the velocity satisfy $\overline{a}_\mathrm{k} = a_\mathrm{k}[1 + \epsilon(\widetilde{G}(k)-1)]$ where $\overline{a}_\mathrm{k}$ are the coefficients of the smoothed field, $\widetilde{G}(k)$ is the Fourier transform of the filter function, and $k$ is the spatial frequency $k = 2\pi / x$.  In the limit $k \rightarrow \infty$, the coefficients satisfy $\overline{a}_\mathrm{k} \rightarrow (1 - \epsilon) a_\mathrm{k}$.  The value of $ \epsilon $ controls the magnitude of the smoothing on a scale of $ \leq \overline{h} $ , and is constrained to $0 < \epsilon \leq 1$.  We choose $\epsilon = 0.8$ to be consistent with the tests in \cite{Monaghan2011}.

We discretise equation~(\ref{eq:xsphsmoothing}) as,

\begin{equation}
\label{eq:smoothdiscrete}
\overline{f}_a = f_a + \epsilon \sum_b \frac{m_b}{\langle\rho_{ab}\rangle_{\single{h}}} (f_b - f_a) W(|\mvec{x_a} - \mvec{x_b}|, \overline{h}_{ab}),
\end{equation}

\noindent where $ f_a $ represents the quantity at particle $ a $, $ \overline{h}_{ab} $ is the arithmetic mean of $ \overline{h}_a $ and $ \overline{h}_b $\footnote{Taken as $ \overline{h}_{ab} = \frac{1}{2}(\overline{h}_a + \overline{h}_b) $ in order to equally weight each smoothing scale.}, $\langle\rho_{ab}\rangle_{\single{h}}$ is the harmonic mean of the densities $\single{\rho}_a$ and $\single{\rho}_b$\footnote{The harmonic mean is of the form $\langle\rho_{ab}\rangle_{\single{h}} = 2\single{\rho}_a\single{\rho}_b / (\single{\rho}_a + \single{\rho}_b)$ and weights toward the lowest value. This allows high density particles to have a fair contribution to the differences within the kernel.}, and the sum is taken over $ b $ nearest neighbours.  We also require doubly-filtered quantities, which involves another application of equation~(\ref{eq:smoothdiscrete}) to the singly-filtered quantities,

\begin{equation}
\label{eq:smoothdiscrete2}
\widehat{\overline{f}}_a = \overline{f}_a + \epsilon \sum_b \frac{m_b}{\langle\rho_{ab}\rangle_{\widehat{h}}} (\overline{f}_b - \overline{f}_a) W(|\mvec{x_a} - \mvec{x_b}|, \widehat{h}_{ab}).
\end{equation}

\noindent In order to calculate the average densities, or weights, in equations~(\ref{eq:smoothdiscrete}) and (\ref{eq:smoothdiscrete2}), we require the density at each particle for a given scale,

\begin{equation}
\label{eq:filterweights}
\begin{split}
\single{\rho}_a = \sum_b m_b W(|\mvec{x}_a - \mvec{x}_b|, \single{h}_{a}), \\
\widehat{\rho}_a = \sum_b m_b W(|\mvec{x}_a - \mvec{x}_b|, \widehat{h}_{a}).
\end{split}
\end{equation}

It is important to note that the values of $\single{S^*}_{ij}$ and $\double{S^*}_{ij}$ are built from the smoothed velocity field and are not smoothed versions of the trace-free symmetric shear tensor in equation~(\ref{eq:symmetricsheargizmo}) \citep{Schmidt2015}.  They have the following corresponding equations,

\begin{equation}
\label{eq:filteredshear}
\begin{split}
\single{S^*}_{ij} = \frac{1}{2}\bigg(\frac{\partial \single{u}_i}{\partial x_j} + \frac{\partial \single{u}_j}{\partial x_i}\bigg) - \frac{1}{3}\delta_{ij}\frac{\partial \single{u}_k}{\partial x_k}, \\
\double{S^*}_{ij} = \frac{1}{2}\bigg(\frac{\partial \double{u}_i}{\partial x_j} + \frac{\partial \double{u}_j}{\partial x_i}\bigg) - \frac{1}{3}\delta_{ij}\frac{\partial \double{u}_k}{\partial x_k}.
\end{split}
\end{equation}

\noindent The magnitudes $|\single{S^*}|$ and $|\double{S^*}|$ are given via equation \ref{eq:magsheargizmo}.

The dynamic method relies on local scale-similarity in the neighbourhood of a point $\mvec{x}$, which in turn implies that the Smagorinsky model is an accurate description of the flow, albeit with a variable constant.  This assumption breaks down in highly complex flows and in some cases, the dynamic model predicts negative values \citep{Piomelli1994, Urzay2013}. Negative values of $C_\mathrm{dyn}$ are usually explained as \textit{backscatter} in the cascade \citep{Piomelii1991, Piomelli1994, Meneveau2000, Vreman2004, Urzay2013} where in some circumstances a fraction of the energy cascading to small scales can return to large scales as  smaller eddies unite to form larger eddies.  An alternate explanation for \smag$<0$ is that the Smagorinsky model fails, and a more appropriate model should be employed.  For now, we adopt the latter view and follow the usual approach in restricting $C_\mathrm{dyn}$ to the range $C_\mathrm{dyn} \in [0, C_s]$ with values $C_\mathrm{dyn}<0$ set to $C_\mathrm{dyn}=0$ \citep{Garnier2009, Schmidt2015}.  The upper limit is imposed since large values are thought to be due to numerical instability.  For the remainder of this paper, we identify and discuss \smag$=\sqrt{C_\mathrm{dyn}}$.  The distinction between the simple and dynamic model is distinguishable based on context.

\begin{table}
 \begin{tabular}{lllll}
  \hline
  Name & Dynamic & Thermal Energy & Velocity & Metals \\                            
  \hline
  None & N/A & \xmark & \xmark & \xmark\\
  S-uz & \xmark & \checkmark & \xmark & \checkmark\\
  D-uz & \checkmark & \checkmark & \xmark & \checkmark\\
  S-uvz & \xmark & \checkmark & \checkmark & \checkmark\\
  D-uvz & \checkmark & \checkmark & \checkmark & \checkmark\\
  \hline
 \end{tabular}
 \caption{We compare five mixing models involving combinations of the standard implementation, dynamic implementation, as well as the mixing of energy, velocity, and metals. We prefix models using the standard implementation by S-, and those involving the dynamic model by D-. Models which mix thermal energy, velocity (momentum), or metals have combinations of the suffixes u, v, or z, respectively.}
 \label{tbl:modeldescription}
\end{table}

\section{Hydrodynamical tests}
\label{sec:hydrotests}

As indicated previously, a number of studies have carried out robust validation of the dynamic Smagorinsky model against, for example, experimental results within the fluid mechanics community \citep{Kleissl2006, Benhamadouche2017, Lee2017, Kara2018, Taghinia2018} and the model has also been adopted by other users, including researchers studying atmospheric phenomena  (e.g. \citealt{Kirkpatrick2006}).   To motivate its use in cosmological and astrophysical simulations, we start by discussing the model within the context of three hydrodynamical tests.

First, we investigate the distribution of predicted \smag values in homogeneous driven turbulence, and the sensitivity of the distributions to variations in the smoothing parameter, $\epsilon$.  Next, we examine the evolution of a Keplerian disc where turbulence is not expected to develop \textit{a priori} yet numerical instabilities and noise lead to disorder in the velocity fields, and subsequent artificially enhanced diffusivities via the trace-free shear strength, $|S^*|$.   Last, we consider the linear regime of the Kelvin-Helmholtz instability which suffers from similar challenges as the Keplerian disc.  In both cases, numerical instability causes the constant-coefficient Smagorinsky model to fail, and we investigate whether the dynamic model can mitigate spurious sub-grid turbulent mixing. 

\subsection{Homogeneous turbulence}
\label{sec:driventurb}

\begin{figure*}
	\centering
	\includegraphics[width=\linewidth]{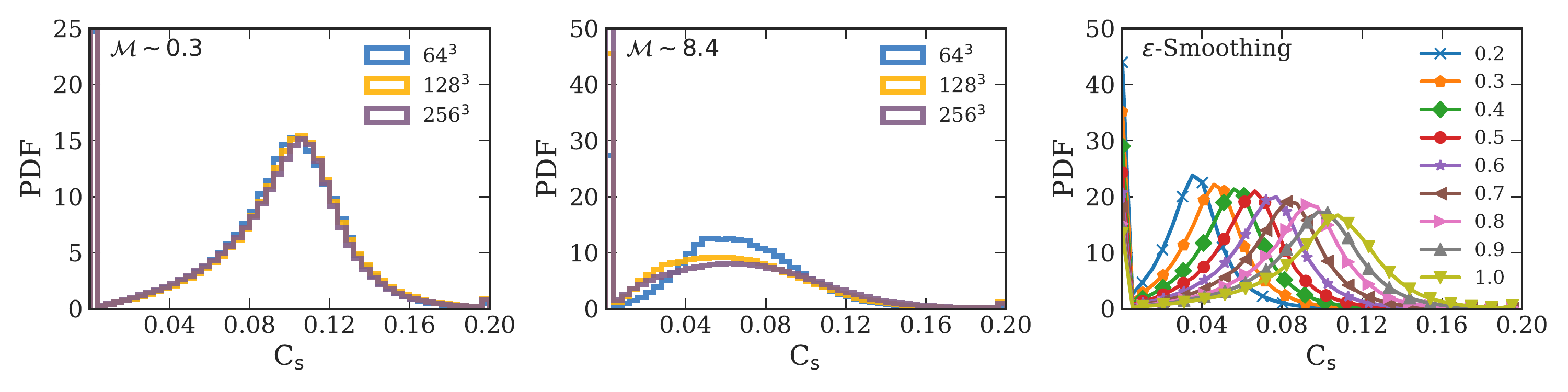}
	\caption{(\textit{left}) Probability density function of \smagend, calculated with the dynamic model, in homogeneous subsonic turbulence at three resolutions --  $64^3$, $128^3$, and $256^3$ particles. The MFM method with a cubic spline kernel is employed, with $N_\mathrm{ngb} = 32$.  Median values of the predicted Smagorinsky constant are ${C}_\mathrm{s, 64} = 0.1005$, ${C}_\mathrm{s, 128} = 0.1017$, ${C}_\mathrm{s, 256} = 0.1009$ and are well within 2\% at maximum difference. (\textit{middle}) Supersonic, homogeneous turbulence at three resolutions, with ${C}_\mathrm{s, 64} = 0.0668$, ${C}_\mathrm{s, 128} = 0.0632$, ${C}_\mathrm{s, 256} = 0.0679$.  (\textit{right}) Probability density function of \smag as we vary the smoothing parameter, $\epsilon$, in homogeneous, subsonic turbulence using the MFM method, with a quintic spline kernel and $N_\mathrm{ngb} = 64$.}
	\label{fig:dynsmagconvergence}
\end{figure*}

We investigate homogeneous, isotropic, driven turbulence to determine to what degree the dynamic model predicts different diffusivities on a per particle basis.

We initialise periodic boxes of side length $L = 1$ with $64^3$, $128^3$, and $256^3$ equal mass particles of an ideal isothermal gas, initial density $\rho = 1$, and energy per unit mass $u = 1000$, placed on uniform Cartesian grids\footnote{Units are arbitrary code units.}.  Following the methods of \cite{Bauer2012} and \cite{Hopkins2015a}, as the system is driven, the thermal energy of the gas particles are reset to the initial value in order to simulate isothermal turbulence.  We investigate the five combinations of mixing models described in Table \ref{tbl:modeldescription}, and examine subsonic ($\mathcal{M} \approx 0.3$) and supersonic ($\mathcal{M} \approx 8.4$) test cases.  

In order to mix the fluid over time, we use an identical forcing routine as in \cite{Bauer2012}.  The accelerations are calculated in Fourier space and only contain power over a small range of modes corresponding to a spatial range $\ell \in [L/2, L]$ (i.e., the largest scales), and the Fourier mode phases are drawn from an Ornstein-Uhlenbeck process.  In the subsonic case, the forcing is purely solenoidal (or incompressible) since the compressive part of the acceleration is removed via a Helmholtz decomposition in Fourier space.  It is important to note that \cite{Grete2018} showed that this is not completely correct, and that compressive modes still exist even with purely solenoidal forcing.  However, we are comparing the effects of the dynamic model using the same forcing methodology across all of our test cases, and additionally we construct the shear and sub-grid scale stress tensor to be trace-free, removing any contributions from compression of the fluid.  We use the exact parameters in Table 1 of \cite{Bauer2012}, and point the interested reader to their Section 2.2 for the precise details of the driving routine. The systems enter an approximate steady state after $t\gtrsim5$.  We measure the probability density functions (PDFs) of \smag in each test in order to determine its sensitivity to the smoothing parameter $\epsilon$ in equation~(\ref{eq:smoothdiscrete}).  In addition, we measure the PDF of the density field in order to gauge the ability of each model to resolve different density ranges in the turbulent flow.

The left panel in Fig. \ref{fig:dynsmagconvergence} shows the distribution of \smag in the subsonic case as predicted with the dynamic model for three separate resolutions: $64^3$, $128^3$, and $256^3$.  The median value and the shape of the distributions do not change much with resolution, indicating excellent convergence.  At $64^3$ resolution we find a median value \smag$=0.1$ and approximately 9.61\% of the particles have \smag$=0$.  In the fluid mechanics literature, as many as 50\% of the fluid elements have been reported to have \smag$=0$ \citep{Piomelii1991, Urzay2013}.  We also test for convergence in supersonic turbulence (see the middle plot in Fig. \ref{fig:dynsmagconvergence}).  Compared to the subsonic case, the dynamic model predicts more particles at \smag$=0$, with a total fraction below 50\%.  The median value is much lower than in the subsonic case,  \smag$=0.066$.  The lower median agrees with calibration results from \cite{Colbrook2017} who found that \smag$\approx0.05$ reproduces the turbulent scaling relationships in supersonic turbulence.

We also investigate the sensitivity of the \smag to the smoothing parameter, $\epsilon$, from equation~(\ref{eq:smoothdiscrete}).  The right plot in Fig. \ref{fig:dynsmagconvergence} shows the PDFs of \smag in homogeneous, subsonic turbulence using the MFM method.  We vary $\epsilon$ between $0.2$ and $1.0$ since $\epsilon$ cannot be greater than $1.0$, as it is derived from a series expansion, and should be $\geqslant 0$ in order to have positive kinetic energies in the smoothed fields \citep{Monaghan2011}.  Values in the range $0.7 \leqslant \epsilon < 1.0$ produce comparable distributions with medians \smag$\approx 0.1$.  \cite{Monaghan2011} found, using a version of SPH with smoothed velocities, that $\epsilon = 0.8$ reproduced turbulent flow trends in decaying wall-bounded turbulence.  For this reason we employ $\epsilon = 0.8$ in all of our tests.

\begin{figure}
	\centering
	\includegraphics[width=\columnwidth]{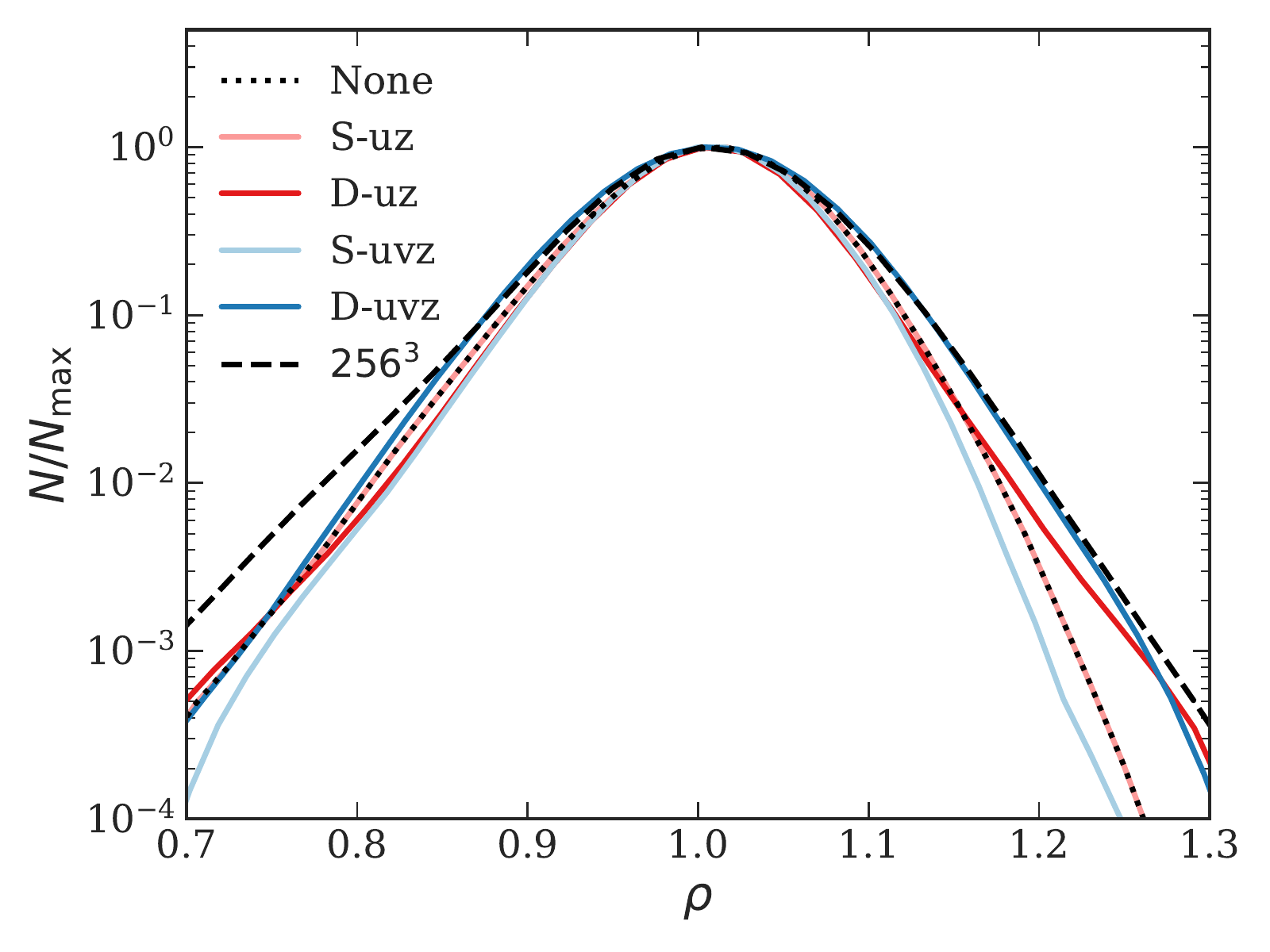}
	\caption{Histograms of the density field in homogeneous subsonic turbulence ($64^3$ case), normalised to the bin with the maximum particle count.  Density measurements taken from $150$ snapshots between $t = 10$ and $t = 25$.  Here, the \model{None} and \model{S-uz} cases coincide.  The particle density contrast in the \model{S-uvz} case is much tighter than the other cases with particles closer to the mean density ($\rho = 1$).  The \model{D-uvz} case is able to represent the widest range of densities, including much higher density regions, with the same number of fluid elements.}
	\label{fig:rhohist}
\end{figure}

Turning to the gas properties, Fig. \ref{fig:rhohist} shows histograms of the gas densities in the $64^3$ homogeneous subsonic turbulence simulations, averaged at $150$ equally spaced times from $t = 10$ to $t = 25$ (inclusive), normalised such that each maximum is $N_\mathrm{max} = 1$.  After $t = 10$, each turbulent field across all models is in an approximate statistically stationary state with $\overline{\rho} \approx 1$.  Although the model labels include the \model{-z} flag, we do not include metal mixing in our driven turbulence simulations.  We retain the suffix for cross-comparison across the different cases in paper, and point the reader to Section \ref{sec:kelvinhelmholtz} for a discussion of turbulent metal mixing in an idealised experiment.  We note that the \model{None} and \model{S-uz} cases coincide and the lines in the figure overlap, indicating that internal energy diffusion with a global \smag value has no effect on the density distribution.

First, using the \model{None} case as a reference curve, the \model{D-uz} case shows a narrower distribution between $0.75 < \rho < 1.15$ with a prominent extended tail toward higher densities.  This indicates that the \model{D-uz} case can represent a wider range of densities in the turbulent cascade.  Since the \model{S-uz} case follows the \model{None} case exactly, this suggests that the localisation of the diffusivity $D$ impacts the density resolution much more than the dependence on velocity shear.

Introducing momentum diffusion alters the density distributions significantly compared to the \model{None} case.  In the \model{S-uvz} case, the density distribution is tighter and exhibits no apparent wings, with the majority of densities falling in the range $0.75 < \rho < 1.2$.  Here the increased diffusivity destroys any small scale structure by causing densities to remain closer to the mean.  However, when we employ the \model{D-uvz} model, we find the opposite effect -- localising momentum diffusion leads to a wider range of densities in the turbulent gas compared to the all other cases.  The effect is strongest at higher densities and therefore we can conclude that the dynamic model is able to resolve higher densities in a turbulent flow at the same mass resolution.

\subsection{Keplerian disc}
\label{sec:kepleriandisc}

In numerical studies of galaxy formation, inherent or artificial dissipation in the hydrodynamical method can cause gas to lose angular momentum and flow radially inward \citep{Hosono2016a}, i.e. the viscous instability.  Numerical simulations require sub-grid diffusive terms as they cannot resolve the viscous scale, however, additional momentum diffusion enhances the viscous instability.  This is an important consideration for the constant-coefficient Smagorinsky model:  in a simulation of a gaseous disc where the rotational velocity curve depends on the radius, including momentum diffusion will trigger the viscous instability even if there is no turbulence as the radial velocity gradient contributes to $|S^*|$.  A quick analytic calculation demonstrates this.  Let us consider a 2D idealised rotating gaseous disc with constant surface density that follows a Keplerian velocity profile $v_\phi \propto r^{-1/2}$.  For this disc, $|S^*| \propto r^{-3/2}$ and inserting this into equation~(\ref{eq:smaggizmo}) with $h$ = const., we find $D \propto r^{-3/2}$.  Generally, for any non-constant velocity profile $v_\phi = v_\phi(r)$, $D \propto \partial_r v_\phi$ in the constant density case.

In a Keplerian disc, particles near the inner radii of a rotating disc will diffuse the strongest in the standard Smagorinsky model as the difference in velocity between each concentric ring is much higher in this region; leading to the rapid break-up of the disc.  One could mitigate the over-diffusion by using a smaller value of \smag in equation~(\ref{eq:smaggizmo}) but then the model would lose its advantages in turbulent flows.  The dynamic model provides a solution to this problem.

We use the 2D idealized Keplerian disc as a representative case of an astrophysical laminar shearing flow to illustrate the aforementioned over-diffusion and compare to the results of the dynamic model.  We simulate a gas annulus of constant surface density using the MFM method, with particles initialised on circular orbits about the centre.  We smooth the inner and outer edges of the annulus in order to suppress numerical instabilities that occur at sharp boundaries.  The particles are subject to an external softened gravitational acceleration ($\mvec{a} = -\mvec{r}(r^2 + \epsilon^2)^{-3/2}$) directed towards the centre of the annulus, and follow a corresponding Keplerian velocity profile.  This initial condition is identical to that in Section 4.2.4 of \cite{Hopkins2015a}, with surface density as a function of radial distance, $r$,

\begin{equation}
    \Sigma(r) =
	\begin{cases}
		(2r)^3 & r < 0.5, \\
		1 & 0.5 \leqslant r \leqslant 2, \\
		(1 + 10(r - 2))^{-3} & r > 2.
	\end{cases}
\label{eq:kepleriandisksurface}
\end{equation}

\begin{figure*}
	\centering
	\includegraphics[width=\linewidth]{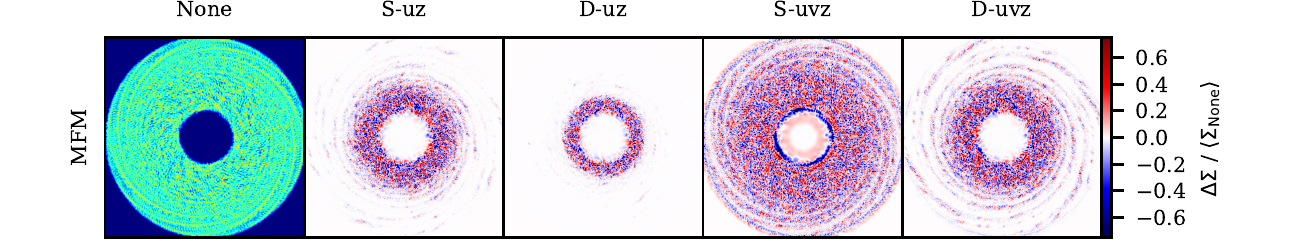}
	\caption{Normalised surface density differences between each mixing model.  The leftmost plot shows the surface density profile of the \model{None} case for comparison.  The \model{S-} models lead to a more rapid break-up of the disc, especially in the case including momentum diffusion (\model{S-uvz}).  The \model{D-uz} model minimises the difference to the \model{None} case.  The dynamic diffusion of thermal energy and momentum (\model{D-uvz}) leads to an equivalent amount of differences to the \model{S-uz} case but the break-up of the disc, due to the over-diffusive \model{S-uvz} case, is avoided.}
	\label{fig:kdmassdiff}
\end{figure*}

In the ideal case, the disc should remain intact at any time $t > 0$.  We study the surface densities of the test cases involving thermal energy and momentum diffusion, described in Table \ref{tbl:modeldescription}.  As with the driven turbulence experiments, we do not include metals despite the model suffix \model{-z}.  We include the suffix to allow for cross-comparison across the various physical tests.  The leftmost plot in Fig. \ref{fig:kdmassdiff} shows the surface density of the disc in the \model{None} model at $t \approx 2 t_\mathrm{orb,r = 1}$.  We focus on relatively early times to decouple the effects of inherent numerical diffusion in the MFM method with those of the turbulent mixing models.  In the \model{None} case, we see that the inner half of the disc is noisy and in the outer region, there are density waves propagating outward, similar to the results in \cite{Hopkins2015a}.  The noise in the inner region, where the orbital time is short, is due to numerical diffusion randomising the particle motions; short of altering the hydrodynamical solver, this effect is unavoidable and is present in all of the tests we investigate here, with or without mixing.   We therefore use the \model{None} model as a baseline experiment to compare the four mixing models.  

In the rest of the four panels in Fig. \ref{fig:kdmassdiff}, we show the point-wise difference in surface density between the model in question, $\Sigma_\mathrm{i}(r)$, and the \model{None} case, $\Sigma_\mathrm{None}(r)$, normalised to the mean surface density in the \model{None} case; i.e. $ \Delta \Sigma(r) / \langle\Sigma_\mathrm{None}\rangle = (\Sigma_\mathrm{i}(r) - \Sigma_\mathrm{None}(r)) / \langle\Sigma_\mathrm{None}\rangle$.  This allows us to compare the diffusion of energy and momentum spatially, by observing the differences directly on the surface of the disc.

In the mixing tests without momentum diffusion, \model{S-uz} (second panel in Fig. \ref{fig:kdmassdiff}) and \model{D-uz} (third panel, Fig. \ref{fig:kdmassdiff}), the inner region ($0.5 < r < 1.0$) of each annulus shows differences compared to the \model{None} case.  These are due to the false identification of turbulence caused by two effects: (i) both models identify random particle motions, like those in the central region, with turbulence and (ii) the diffusivity scales as $D \propto r^{-3/2}$ in the \model{S-uz} case.  The advantage of the dynamic model is that the radial extent of the affected region is significantly smaller compared to the \model{S-uz} case.  \model{D-uz} predicts much smaller values of \smag (median of $C_\mathrm{s} \approx 0.026$ in the \model{D-uz} case compared to \smag$ = 0.2$ in \model{S-uz}) and, therefore, the differences with the \model{None} case are mostly limited to the relatively noisy central annulus.

\begin{figure}
	\centering
	\includegraphics[width=\linewidth]{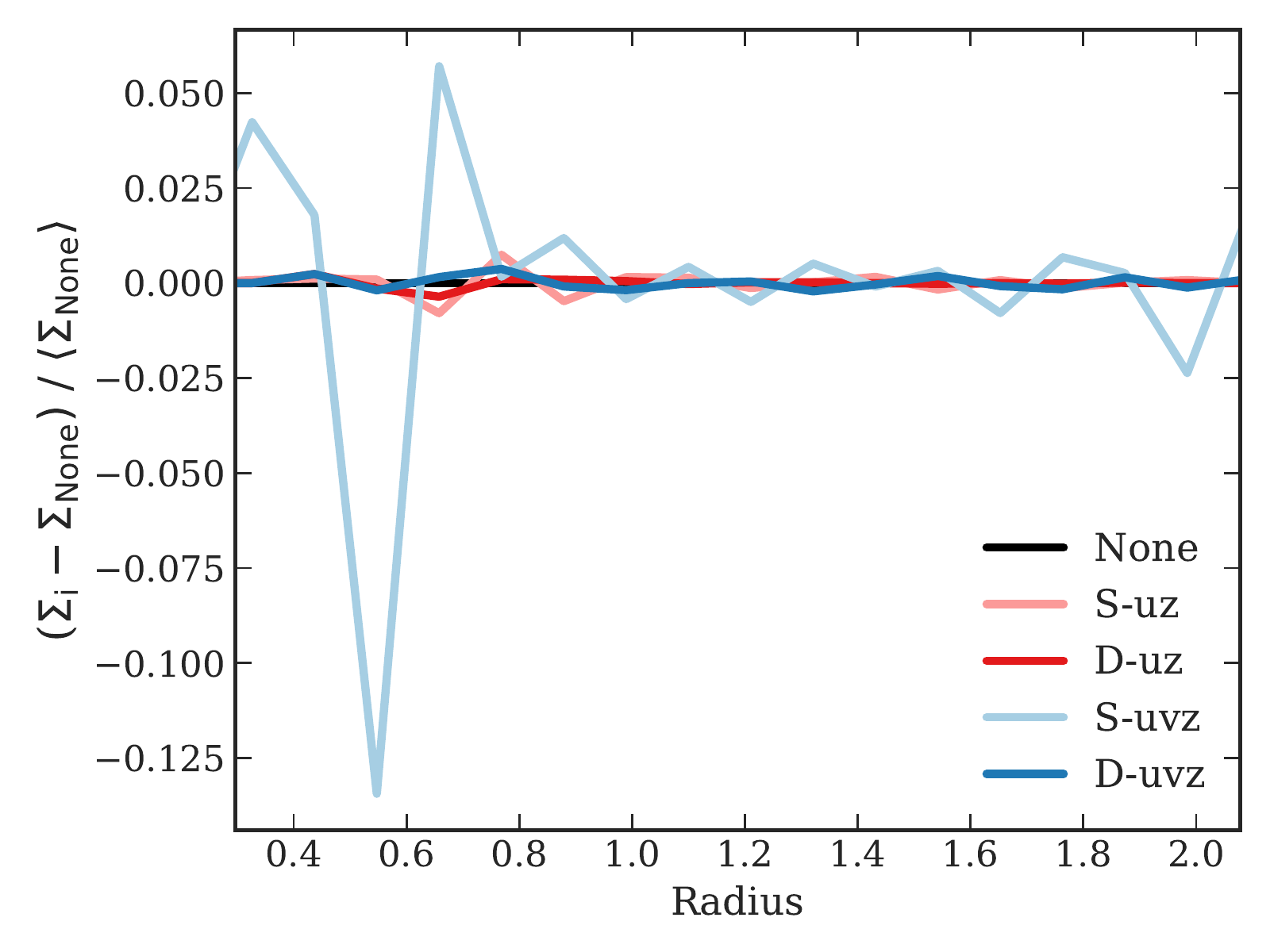}
	\caption{The azimuthally averaged differences in surface density for the Keplerian disc experiments, between each mixing model $i$ and the \model{None} case at $t \approx 2 t_\mathrm{orb} $.  The \model{D-uz}, \model{D-uvz}, and \model{S-uz} cases show small fluctuations around the \model{None} case whereas the \model{S-uvz} model causes large differences due to over-diffusing momentum.}
	\label{fig:kdsurfprof}
\end{figure}

From Fig. \ref{fig:kdmassdiff} we see that the addition of momentum diffusion (the \model{S-uvz} and \model{D-uvz} cases, two rightmost panels) causes increased noise throughout the disc.  Specifically for the \model{S-uvz} case, the noise in the disc extends radially to the outer boundary, and the material spreads into the central region, compared to the \model{None} case (notice the faint pink annulus).  There is also a corresponding deficit of gas at $R \approx 0.5$ (blue ring) showing that gas has collapsed due to the viscous instability.  Comparing \model{D-uvz} to the \model{None} case, we see that the differences do not extend to the boundary, but rather follow the density waves in the disc caused by natural dissipation in MFM.  The central noisy region has the same extent as in the \model{S-uz} case but with no apparent in-fall of material into the central region.  These results indicate that the dynamic model allows momentum diffusion in laminar shear flows without instigating the viscous instability.

In Fig. \ref{fig:kdsurfprof}, we show the azimuthally averaged difference between the surface density in each mixing model, $\Sigma_\mathrm{i}$, and the \model{None} case, $\Sigma_\mathrm{None}$, in order to get a more quantitative estimate of the extent of over-diffusion.  Both \model{S-uz} and \model{D-uz} are nearly identical to the \model{None} case, except in the innermost region ($R \lesssim 1$ for \model{S-uz} and $R \lesssim 0.7$ for \model{D-uz}) where there are slight fluctuations about the \model{None} value.  The differences compared to the \model{None} case are reduced because of the azimuthal averaging.

\begin{figure*}
	\centering
	\includegraphics[width=\linewidth]{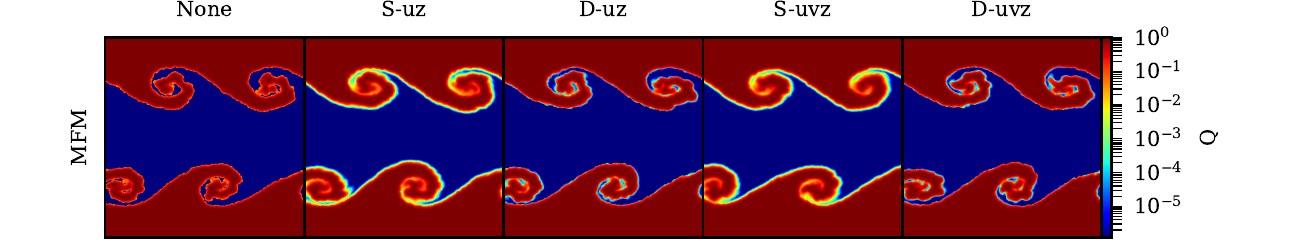}
	\caption{A comparison of tracer concentrations $Q$, in the 2D Kelvin-Helmholtz instability test at $t = 0.28 \tau_\mathrm{KH}$, simulated with the MFM method.  The columns represent the five turbulent mixing models.  In the \model{None} case, gas particles cannot exchange the tracer concentrations and the interface of the instability remain unsmoothed.  In this case, inter-fluid mixing only occurs when particles move across the boundary. The \model{S-} cases diffuse rapidly due to the presence of strong shear at the boundary, and the tracer engulfs the whorls during the early evolution of the instability.  The \model{D-} cases provide a compromise between the two situations -- they limit diffusion strictly to the interface between the two fluids, and the internal structure of the whorls are distinguishable.}
	\label{fig:khtracerearly}
\end{figure*}

Introducing momentum diffusion results in greater quantitative differences in the disc.  The \model{S-uvz} model results in significant flows (inward and outward) in the annulus relative to the \model{None} case.  For example, angular momentum transport causes some of the material at $R \approx 0.5 - 0.6$ to flow inward due to loss of angular momentum, and some to flow outward.  This occurs throughout the disc, resulting in fluctuations extending to the outer edge of the disc.  These flows lead to the \model{S-uvz} case showing regions of higher densities between $0.6 \lesssim R \lesssim 1.8$.  Comparatively, the small fluctuations in the \model{D-uvz} case are similar to the models without momentum diffusion.  The dynamic model clearly reduces the impact of the viscous instability, but a question then arises: why does the dynamic model not eliminate the viscous instability completely?

The dynamic model formally predicts \smag$=0$ for all of the gas particles.  Prior to $t \approx 0.4 t_\mathrm{orb}$, virtually all of the particles have a near zero value of \smagend.  However, by $t \approx 2 t_\mathrm{orb} $, turbulent momentum diffusion and the inherent noise in the inner regions lead to a small non-zero distribution centred at approximately \smag$\approx 0.04$ (ignoring particles with \smag$=0$).  Overall, the particles have a median value of \smag$\approx 0.026$ when including the 21.4\% of particles at \smag$=0$.  Although we cannot avoid the inherent numerical noise, due to MFM's Riemann solver, the dynamic method does minimise the damage: in standard implementations, \smag is in the range $\approx 0.1 - 0.2$ \citep{Garnier2009} while in the dynamic model, only a small fraction ($\approx 7$\%) of the particles attain such values. 

These tests show that in the case of a rotating, laminar shear flow, the dynamic model (\model{D-uvz}) can indeed minimise turbulent mixing of thermal energy and momentum -- preventing unphysical viscous flows within the disc.  Therefore, we recommend incorporating the dynamic Smagorinsky model in all numerical simulations involving rotating galactic discs which simultaneously include turbulent mixing models.

\subsection{The Kelvin-Helmholtz instability}
\label{sec:kelvinhelmholtz}

In a fluid with high velocity shear, or at the shearing interface between two fluids, rapidly growing perturbations cause mixing within the fluid.  In the case of a shear interface, the perturbations cause the two fluids to encroach the boundary, transporting and mixing fluid properties such as thermal energy, momentum, and metals.  This is the Kelvin-Helmholtz (KH) instability, and the timescale characterising the growth of perturbations is given by,

\begin{equation}
\label{sec:khtimescale}
\tau_\mathrm{KH} = \frac{(\rho_1 + \rho_2)}{\sqrt{\rho_1\rho_2}}\frac{\lambda}{\Delta v},
\end{equation}

\noindent where $\rho_1$, $\rho_2$ are the densities of the two fluids, $\Delta v$ is the velocity difference, and $\lambda$ is the wavelength of the perturbation \citep{Chandrasekhar2013}.  In the $t < \tau_\mathrm{KH}$ regime (the linear regime), the flow has not completely transitioned to turbulence and sub-grid turbulent mixing does not dominate the resolved mixing.

Shear flows, and hence KH instabilities, are ubiquitous in galactic environments: ram pressure stripping of galaxies falling into groups and clusters, galaxy mergers, galactic winds streaming into the circumgalactic medium, and a myriad of other processes involve the KH instability.  The constant-coefficient Smagorinsky model over-diffuses in such situations because it identifies shearing motion with turbulence, as discussed in Section \ref{sec:kepleriandisc}, and the presence of high shear increases the diffusivity to the maximum at the interface.  Ideally, sub-grid turbulent mixing models that better reflect physical reality are preferable. More precisely, models that capture unresolved mixing in turbulent situations and avoid diffusion in laminar shear flows.

We investigate a simple 2D KH test in order to demonstrate the over-diffusive nature of the standard Smagorinsky model and determine if the dynamic model mitigates the problem.  We set up a 2D configuration of $256^2$ ideal gas particles in a square ($L = 1$) domain, with constant pressure, and with an initial density profile,

\begin{equation}
\label{eq:khdensity}
\rho(y) =
\begin{cases}
\rho_2 - (\Delta \rho / 2)\exp{[(y - 1/4)/\Delta y]}, & 0\leqslant y < 1/4\\
\rho_1 + (\Delta \rho / 2)\exp{[(1/4 - y)/\Delta y]}, & 1/4 \leqslant y < 1/2\\
\rho_1 + (\Delta \rho / 2) \exp{[(y - 3/4)/\Delta y]}, & 1/2 \leqslant y < 3/4\\
\rho_2 - (\Delta \rho / 2) \exp{[(3/4 - y)/\Delta y]}, & 3/4 \leqslant y < 1\\
\end{cases}
\end{equation}

\noindent and initial velocity profile,

\begin{equation}
\label{eq:khvelocity}
v_x(y) =
\begin{cases}
-1/2 + (1/2) \exp{[(y - 1/4)/\Delta y]}, & 0\leqslant y < 1/4\\
1/2 - (1/2) \exp{[(1/4 - y)/\Delta y]}, & 1/4 \leqslant y < 1/2\\
1/2 - (1/2) \exp{[(y - 3/4)/\Delta y]}, & 1/2 \leqslant y < 3/4\\
-1/2 + (1/2) \exp{[(3/4 - y)/\Delta y]}, & 3/4 \leqslant y < 1.\\
\end{cases}
\end{equation}

\noindent We choose $\rho_2 = 2$, $\rho_1 = 1$, and $\Delta y = 0.025$, and introduce a sine wave velocity perturbation, with period $T = 2$ and amplitude $A = 0.01$, at $t = 0$.  This gives a perturbation wavelength $\lambda = 1/2$, and therefore $\tau_\mathrm{KH} \approx 0.71$.  In addition, we add a uniform passive scalar tracer of concentration $Q = 1$ to all gas particles in the range $0 \leqslant y < 1/4$ and $3/4 \leqslant y < 1$, which is the higher density gas.  We focus our analysis on the evolution of the tracer  concentration $Q$.

Fig. \ref{fig:khtracerearly} shows the tracer concentration at $t = 0.28 \tau_\mathrm{KH}$ (i.e. in the linear regime).  The five panels show the five mixing models described in Table \ref{tbl:modeldescription}.  We first consider the \model{None} case.  The tracer concentration $Q$ follows the high density regions of the experiment and we see individual particles advecting across the shear interface.  Although there appears to be less tracer on this interface (orange-red line), it is impossible for particles to exchange tracer in the \model{None} case.  The interpolation method we employ causes this effect as it smooths the particle properties over the resolution scale $h$, leading to a minuscule amount of artificial mixing.

When we allow for the turbulent mixing of the tracer in the \model{S-} and \model{D-} cases, we see diffusion occurring along the shear interface.  In both \model{S-} models, $Q \sim 10^{-2}$ at the interface and the tracer engulfs the initially pristine, lower density gas in the whorls.  This is in contrast to the two \model{D-} cases where the majority of diffusion occurs in the whorls themselves with comparatively little along the rest of the interface.  The constant-coefficient Smagorinsky model diffuses the most because of the false identification of strong turbulence through the norm of the shear tensor $|S^*|$.  The velocity profile in equation~(\ref{eq:khvelocity}) shows that although the flow is laminar, there is a gradient, $|\partial v_x / \partial y| > 0$, across the entire domain.  Based on our arguments for the Keplerian disc in Section \ref{sec:kepleriandisc}, the fact that $D \propto |\partial v_x / \partial y|$ directly leads to the over-diffusion of the tracer.  The diffusion coefficient $D$ also depends on $|S^*|$ in the \model{D-} models but in the dynamic model most of the values of \smag are near zero.  Only in the whorls do we find values of \smag as large as \smag$=0.2$ but these are limited to this region -- where the transition to turbulence begins.

Over-diffusion in non-turbulent shear flows can have important consequences for the gas in numerical studies of galaxy formation.  If the KH timescale is longer than the gas cooling time, then the gas is susceptible to over-cooling in the whorls due to metals transferred to the region.  This does not accurately capture what physically occurs at the sub-grid level; in the linear regime, the fluids do not mix completely.  The dynamic model solves this issue by minimising mixing along the interfaces while allowing it to proceed in the whorls where, in principle, the KH instability continues down to unresolved scales.  

\section{Isolated disc galaxy}
\label{sec:isogalaxy}

Here we investigate an isolated galaxy in order to test the effects of localised diffusion in a more realistic, physical environment. 

\subsection{Initial conditions}
\label{sec:isogalaxyics}

We follow the method outlined in \cite{Springel2005b} to set up the initial conditions, using the \pkg{GALSTEP} package\footnote{\url{https://github.com/ruggiero/galstep}} \citep{Ruggiero2016} in combination with \pkg{DICE}\footnote{\url{https://bitbucket.org/vperret/dice}} \citep{Perret2014}.  The galaxy is a Milky Way-like system \citep{Sokolowska2016} consisting of a dark matter halo of mass $M_\mathrm{h} = 10^{12}$ $\Msun$, a gaseous halo of mass $M_\mathrm{gh} = 3\times 10^{10}$ $\Msun$, a stellar bulge of mass $M_\mathrm{b} = 10^{10}$ $\Msun$, and gas and stellar discs of masses $M_\mathrm{g} = 10^{10}$ $\Msun$ and $M_\mathrm{s} = 5\times 10^{10}$ $\Msun$, respectively.  The dark matter and bulge components follow a Hernquist density profile with scale factors $a = 47$ kpc and $a = 1.5$ kpc, respectively.  The stellar and gaseous discs follow an exponential density profile with a radial scale $R_\mathrm{d} = 3.5$ kpc, and the scale-heights for these components are $z_\mathrm{0} = 0.7$ kpc and $z_\mathrm{0} = 0.0175$ kpc, respectively.  We initialize the gaseous halo metallicity at $Z_\mathrm{gh} = 10^{-3} Z_\mathrm{\odot}$, and the gaseous disc metallicity at $Z_\mathrm{g} = Z_\mathrm{\odot} / 3$, with $Z_\mathrm{\odot} = 0.02$ \citep{Anders1989}.  Our fiducial run is carried out at a gas mass resolution of $M_\mathrm{g,res} = 5\times 10^3$ $\Msun$, along with the softening values specified in Table \ref{tbl:isodiskparams}.  We evolve the disc for 2 Gyr in an isolated (non-cosmological) setting.

\begin{table}
 \caption{Parameters for the isolated galaxy.}
 \label{tbl:isodiskparams}
 \begin{tabular}{llll}
  \hline
  Component & Particle Mass (M$_\odot$) & Min. Softening (pc) & N$_\mathrm{part}$ \\                            
  \hline
  Gas & $5.0\times10^3$ & $1.4$ & $8\times10^6$ \\
  Halo & $5.0\times10^5$ & $12.0$ & $2\times10^6$ \\
  Disc & $5.0\times10^5$ & $3.2$ & $1\times10^5$ \\
  Bulge & $2.5\times10^5$ & $1.4$ & $4\times10^4$ \\
  \hline
 \end{tabular}
\end{table}

\subsection{Galactic physics}
\label{sec:isogalaxymodels}

\subsubsection{Star formation \& cooling}
\label{sec:isogalaxysf}

We employ the sub-grid multiphase ISM model of \cite{Springel2003a}, which places the ISM gas ($n_\mathrm{H} > n_\mathrm{*,crit}$, where $n_\mathrm{*,crit}$ is the star formation density threshold) on an effective equation of state (EoS).  This model has been used extensively in numerical galaxy formation studies (for recent examples, see \citealp{Genel2014, Schaye2014, Grand2017}), and provides well-converged results \citep{Springel2005b}.  For primordial and metal-line cooling, we use the \pkg{GRACKLE-2.1} cooling library \citep{Smith2017} in combination with the UV background from \cite{Faucher2009}.

In this sub-grid model, stars form stochastically, on a star formation timescale $t_\mathrm{sfr}$, from gas that reaches a density $n_\mathrm{*,crit}$.  Here we take $n_\mathrm{*, crit} = 0.1$ cm$^{-3}$ and $t_\mathrm{sfr} = 2.1$ Gyr, which give a good fit to the Kennicutt law \citep{Kennicutt1998, Springel2003a}.  We differ from the original model in the choice of the initial mass function (IMF).  We assume the Chabrier IMF \citep{Chabrier2003} instead of the Salpeter IMF \citep{Salpeter1955}. 

\subsubsection{Feedback}
\label{sec:isogalaxyfeedback}

Due to lack of resolution, it is necessary to include a sub-grid model for feedback from stars, including the effects of supernovae, stellar radiation, and stellar winds.  

For massive stars, we adopt the scheme used in the MUFASA simulations \citep{Dave2016c, Dave2017} and refer the reader to these references for a detailed description.  In brief, stellar feedback is expected to drive galactic outflows and, in the MUFASA approach, stellar feedback directly launches a kinetic wind via a two-parameter model that characterises the net effect of stellar feedback into a mass loading factor $\eta$, and the wind speed $v_\mathrm{w}$.  These parameters are calibrated to the FIRE wind scalings \citep{Muratov2015}, where $\eta$ scales with the stellar mass of the host galaxy and $v_\mathrm{w}$ scales with the galaxy's circular velocity.  We fix $\eta$ and $v_\mathrm{w}$ to the values for our isolated system based on equations~(6) and (7) in \cite{Dave2016c}.  At launch, the outflow hydro-dynamically decouples, and only recouples if the wind speed drops to 50\% of the local sound speed, the density of the ambient medium is 1\% of the interstellar medium density, or the outflow has travelled for 2\% of the Hubble time at launch \citep{Dave2016c}. For a more detailed description of the decoupled outflow model, see \cite{Springel2005b}, \cite{Oppenheimer2008}, \cite{Liang2016}, and \cite{Dave2016c}.

\begin{figure*}
	\centering
	\includegraphics[width=\linewidth]{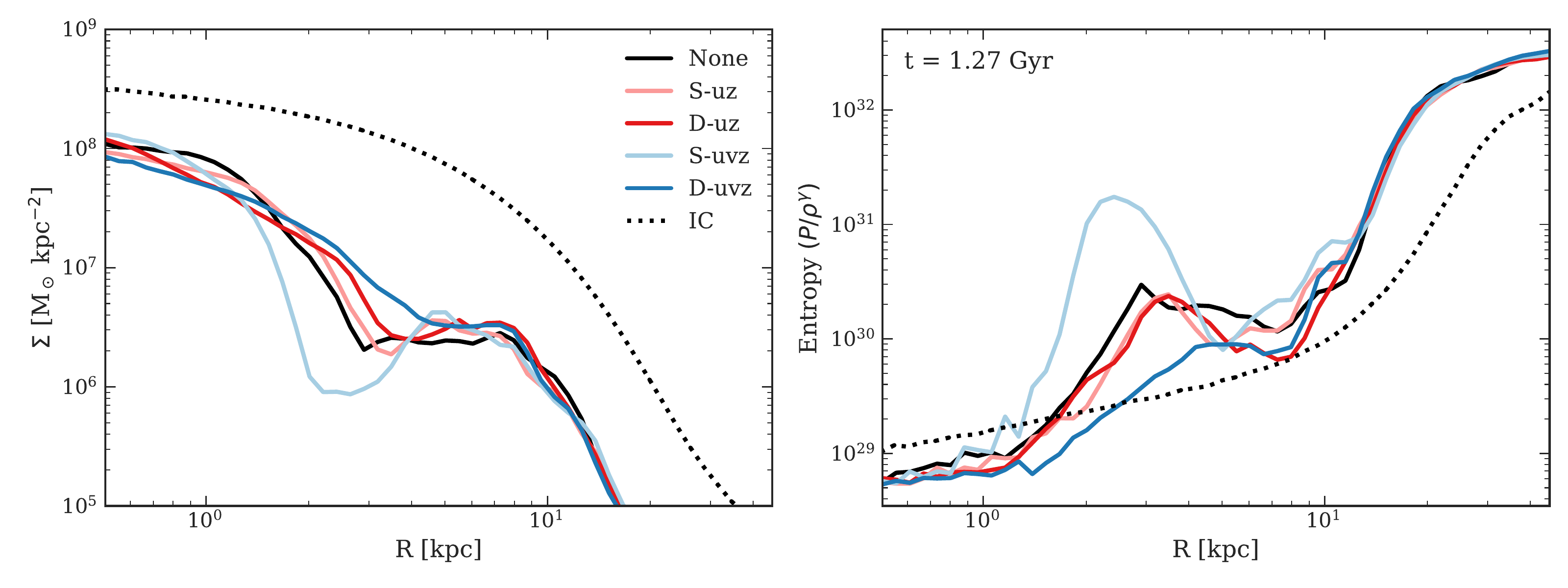}
	\caption{Isolated galaxy radial disc profiles at $t = 1.27$ Gyr, averaged azimuthally and vertically between $\pm0.5$ kpc from the plane of the disc. (\textit{left}) The gas surface density. (\textit{right}) A monotonic measure of the gas entropy.  The dotted line in both plots represents the corresponding profile in the initial condition, which is identical for all models.  Over time, the normalisation of the surface density decreases due to gas consumption and stellar feedback.  The dynamic model with thermal energy and momentum diffusion (\model{D-uvz}) produces a more stable disc whereas the constant-coefficient model (\model{S-uvz}) shows a more concentrated central region.}
	\label{fig:mfm_iso_radial}
\end{figure*}

The contribution from supernovae type Ia (SNIa) are modelled following \cite{Scannapieco2005} as a prompt and delayed component, where the prompt component occurs simultaneously with SNII, and the delayed component begins $0.7$ Gyr after the star formation time.  The prompt component is assumed to release $10^{51}$ erg of thermal energy to the star-forming gas, whereas the delayed component is added in a kernel-weighted manner to the nearest $16$ gas particles. 

\subsubsection{Chemical enrichment}
\label{sec:isogalaxyenrich}

The chemical enrichment of gas is paramount to the study of turbulent mixing as the metallicity follows the mixing of energy and momentum, and, therefore, provides a tracer for the diffusion equation.  We track $11$ chemical elements in our isolated and cosmological simulations: H, He, C, N, O, Ne, Mg, Si, S, Ca, and Fe.  These elements are produced from three sources in the simulations: SNIa, SNII, and the winds from AGB stars.

For SNIa, the prompt component returns mass to the ISM and enriches the star-forming gas instantaneously.  Each SNIa is assumed to release $1.4\Msun$ of metals, with yields from \cite{Iwamoto1999}. For the delayed component, stars deposit metals over their nearest $16$ neighbouring gas particles in a kernel-weighted fashion.

SNII return mass and enrich the gas via the instantaneous recycling approximation \citep{Springel2003a, Oppenheimer2008, Dave2016c} following,

\begin{equation}
\Delta Z_\mathrm{i} = (1 - f_\mathrm{SN})\cdot y_\mathrm{i}(Z)\cdot\frac{\Delta t}{t_\mathrm{sfr}}
\end{equation}

\noindent where $f_\mathrm{SN}$ is the fraction of stars in the Chabrier IMF expected to go supernova, $y_\mathrm{i}(Z)$ is the metallicity dependent yield of species $i$, $\Delta t$ is the timestep, and $t_\mathrm{sfr}$ is the aforementioned star-formation time-scale.  The SNII yields follow \cite{Nomoto2006} and are a function of the metallicity of the gas receiving the metals.  Following \cite{Dave2016c}, the SNII yields are reduced by a factor of $0.5$ in order to match the mass-metallicity relationship.  SNII also return mass into the gas via the instantaneous recycling approximation.

For AGB stars, chemical enrichment is done in a kernel-weighted fashion over the nearest $16$ neighbours.  AGB yields are obtained from a lookup table as a function of age and metallicity based on the study in \cite{Oppenheimer2008}.  The mass-loss rates of the AGB stars are calculated from a lookup table based on \cite{Bruzual2003} stellar models.

\subsection{Results: Disc stability}
\label{sec:isogalaxystability}

Fig.~\ref{fig:mfm_iso_radial} shows radial disc profiles of the surface density (left panel) and entropy\footnote{The entropy scales as $S \propto$ ln$(P / \rho^{\gamma}$), and therefore our measure is off by a multiplicative constant.} (right panel) in our isolated galaxies after $1.27$ Gyr of evolution.  The dotted line represents the initial condition (IC) for all models.  The differences between models only appear after $1$ Gyr ($\approx 4$ rotations in the mid-disc) and continue until star formation consumes the bulk of the gas after $2$ Gyr.

First, we consider the \model{None} case.  The radial surface density (left panel, Fig. \ref{fig:mfm_iso_radial}) gives a measure of the stability of the disc.  We see that by $t = 1.27$ Gyr, compared to the shape of the IC density profile, the gas has moved inward toward the centre, especially within $R < 3$ kpc.  A combination of the bar instability and inherent numerical dissipation causes the gas to concentrate inside $R \approx 2$ kpc, whereas gas consumption and galactic winds due to supernova feedback cause the difference in normalisation compared to the IC.  Correspondingly in the right panel, there is an order-of-magnitude increase in entropy from $R = 1$ kpc to $R = 3$ kpc.

In the \model{S-uz} and \model{D-uz} models, we see only minor difference in the surface density and entropy compared to the \model{None} case.  Evidently, thermal energy diffusion, combined with metal diffusion, has negligible impact on the structure of the disc.

When we introduce momentum diffusion in the \model{S-uvz} and \model{D-uvz} cases, the differences compared to the \model{None} case at $t = 1.27$ Gyr are more significant.  Fig.~\ref{fig:mfm_iso_radial} shows that in the \model{S-uvz} case there is an order-of-magnitude deficit of gas surface density (left plot) between $R = 1$ kpc and $R = 3$ kpc with a corresponding jump in entropy (right plot).  Over-diffusion of momentum due to the diffusivity scaling strongly with the shear causes the inward flowing gas to be more centrally concentrated compared to the \model{None} case, and also engenders an outward flow leading to slightly higher density (again, compared to \model{None}) at $R \approx 4 - 7$ kpc.  Once the instability occurs, the effect accelerates and the trend remains throughout the evolution of the disc.  However, with the dynamic model (\model{D-uvz}), less gas gets redistributed.  From the surface density, we see that the disc stabilises when momentum is diffused locally, based on the turbulent character of the flow.

\subsection{Results: Metal distribution functions}
\label{sec:isogalaxydists}

\begin{figure*}
	\centering
	\includegraphics[width=\linewidth]{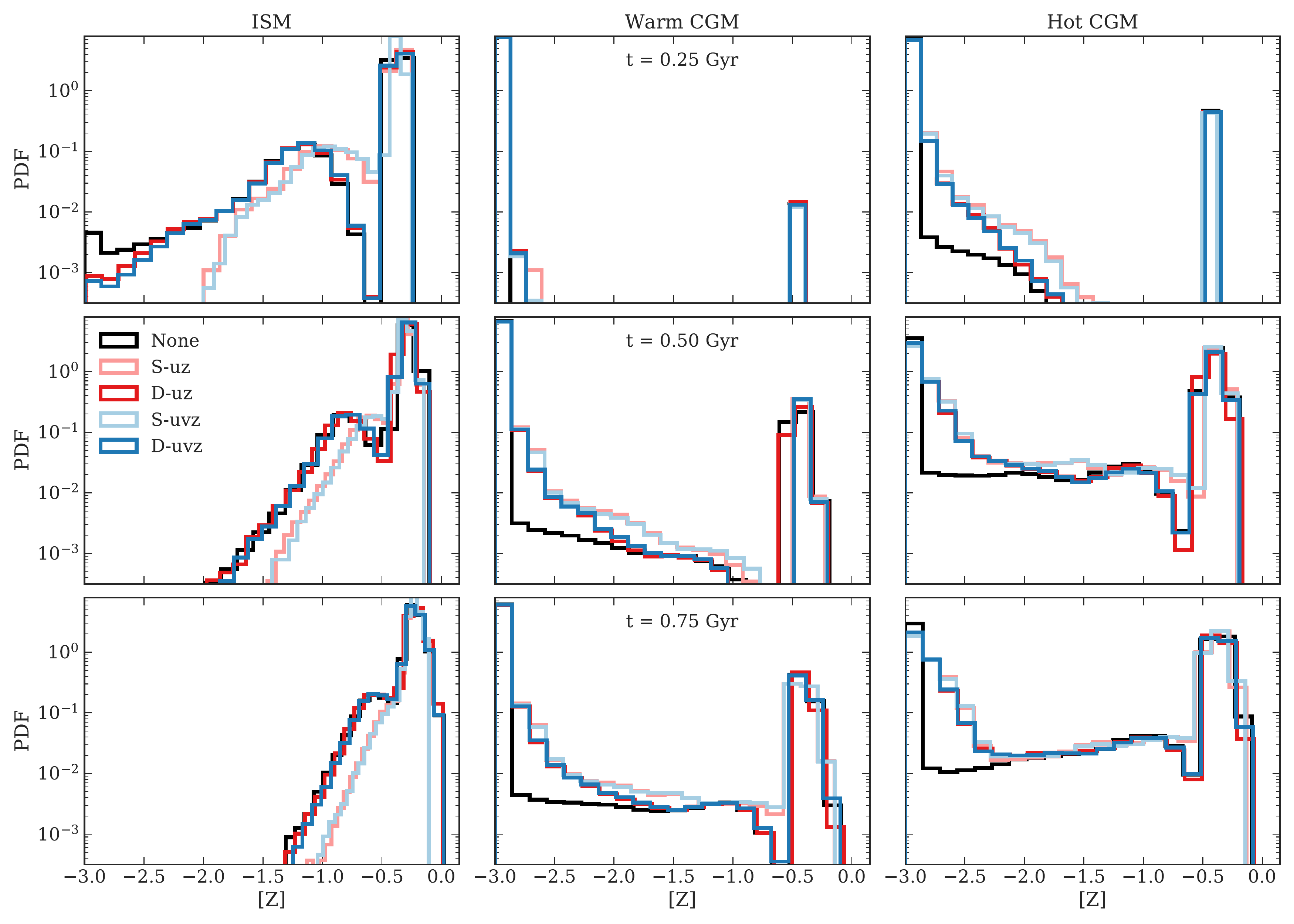}
	\caption{The metal distribution functions (MDFs) for the isolated disc galaxy in solar units ($Z_\mathrm{\odot} = 0.02$).  Each column represents a specific region of $T-\rho$ phase space.  The leftmost column represents the interstellar medium, whereas the middle and rightmost columns show the MDFs for the warm ($10^5$ K $ < T < 10^6$ K) and hot ($T > 10^6$ K) circumgalactic medium, respectively.  Each row represents a different time in Gyr: $t = 0.25$ Gyr, $0.50$ Gyr, and $0.75$ Gyr from top to bottom, respectively.  Dynamic diffusion slows metal mixing in the interstellar medium and causes increased mixing in the circumgalactic medium, at approximately the same levels as the constant-coefficient Smagorinsky model.}
	\label{fig:mfm_iso_metal}
\end{figure*}

Supernovae and stellar winds inject energy and metals into the interstellar medium (ISM), engendering turbulent motion that mixes and spreads thermal energy, momentum, and metals throughout the medium.  These processes also drive a galactic-scale wind that deposits metals and drives turbulence in circumgalactic medium (CGM) \citep{Evoli2011}. 

In order to obtain a measure of energy and metal mixing in the five models under consideration (see Table \ref{tbl:modeldescription}), we examine metal distribution functions (MDFs) in three phases of gas: (1) gas with density above the star formation critical density ($n_\mathrm{H} > 0.1$ cm$^{-3}$) -- i.e, the ISM gas, (2) warm CGM gas in the halo in the range $10^5$K $ < T < 10^6$K, and (3) hot CGM gas with $T > 10^6$K.  The gas density in the latter two phases is $n_\mathrm{H} \leqslant 0.1$ cm$^{-3}$.  In our isolated system, we find that the cool, non-star-forming gas ($T < 10^4$ K) is mostly at the outskirts of the halo and that there are no significant variation in its MDFs between models.  Therefore, we do not discuss this phase further in the isolated case.  Fig. \ref{fig:mfm_iso_metal} shows the MDFs for the the ISM, the warm CGM, and the hot CGM, at three separate evolutionary times: $t = 0.25$ Gyr (top row), $0.50$ Gyr (middle row), and $0.75$ Gyr (bottom row) after the initial conditions for the MFM method.  Henceforth, we use $[Z] \equiv \log{(Z/Z_\mathrm{\odot})}$ as a proxy for metallicity.

First, we focus on the ISM shown in the left column of  Fig.~\ref{fig:mfm_iso_metal}.  At $t = 0.25$ Gyr, the MDF of the \model{None} case is bimodal with a narrow peak at $[Z] \approx -0.3$, a broad distribution at $[Z] < -0.5$ with a peak at $[Z] = -1.2$, and a dearth at $[Z] = -0.5$.  Gas in the ISM with metallicity lower than the initial $[Z] \approx -0.5$ comes from the lower metallicity (initial $[Z] = -3.0$) halo gas that cools onto the disc and is steadily enriched.  Recall that in the \model{None} case, gas cannot exchange metals between particles and therefore it gives an upper bound on the mixing timescale for a given metal injection rate from stars.  The \model{D-} cases follow the \model{None} case with minimal differences, therefore we conclude that the \model{D-} cases provide minimal turbulent mixing in the ISM.  The \model{S-} cases share the basic shape as the other cases but the lower metallicity component is narrower (there is very little gas with $[Z] < -2.0$), and is shifted to the right, with peak $[Z] \approx -0.9$.  The over-diffusivity of the \model{S-} cases is apparent and is caused by the high levels of shearing motion in the supersonic ISM, leading to high values of the diffusivity $D$, similar to the Keplerian disc in Section \ref{sec:kepleriandisc}.

At $t = 0.5$ Gyr in the ISM (left column, middle row), the low metallicity component of the MDF for all cases has narrowed and shifted toward higher metallicity.  In the \model{None} case, the tightening is due to stars depositing metals into the medium and driving all of the gas toward highly metallicities.  Here, the \model{D-} cases follow the \model{None} case closely -- as at $t = 0.25$ Gyr -- while the \model{S-} cases have a smaller spread.   We attribute the differences between the \model{D-} and \model{S-} cases to the dynamic model predicting median diffusivities orders-of-magnitude lower than the constant-coefficient Smagorinsky model in the ISM at $t > 0.25$ Gyr. 

By $t = 0.75$ Gyr, all cases have tight distributions with very few particles having $[Z] < -1.5$.  \model{None} and \model{D-} cases continue to exhibit very similar distributions with a larger spread compared to the \model{S-} cases.

Now we consider the MDFs of the warm CGM, in the middle column of Fig. \ref{fig:mfm_iso_metal}.  In the \model{None} case at $t = 0.25$ Gyr, the distribution is bimodal with peaks at $[Z] \approx -0.5$ and $[Z] = -3.0$.  The peak at $[Z] = -3.0$ is the initial halo metallicity whereas the peak at $[Z] \approx -0.5$ is due to enriched gas either pushed out of the ISM by stellar winds or at the interface between the CGM and ISM.  The \model{D-} and \model{S-} cases closely follow the \model{None} case, but with a slight spread toward $[Z] > -3.0$ for the low metallicity gas.

Next we consider the $t = 0.5$ Gyr and $t = 0.75$ Gyr cases.  In all of the experiments, the flow of the enriched ISM gas into the CGM results in an increase in the fraction of particles near $[Z] \approx -0.5$.  The differences near $[Z] = -3.0$ between the \model{None} and \model{S-}/\model{D-} cases are due to the inability of the gas in former case to mix metals.  For the \model{None} case, a spread only occurs if the gas enters the CGM at $[Z] > -3.0$ from the ISM, or the CGM gas at the ISM-CGM boundary becomes enriched via the kernel-weighting approach associated with delayed SNIa and AGB wind feedback.  Recall that in this implementation, metals are deposited in a kernel-weighted fashion over the nearest $16$ neighbouring particles.  Therefore, gas classified as CGM, that is spatially adjacent to the star-forming gas in the ISM, can be enriched at low levels.  The gas in the turbulent mixing cases consistently exchange metals if the value of \smag is non-zero, and the gas-phase metals in the halo further mix -- leading to the greater spread in the lower peak at $[Z] = -3.0$, corresponding to the bulk of the initial gas.  The \model{D-} cases show the aforementioned spread in the metallicity in the range $-3.0 < [Z] < -1.75$, indicating increased mixing due to sub-grid turbulence, but then coincides with the \model{None} case for metallicities $[Z] > -2.0$.  The \model{S-} and \model{D-} cases share a similar distribution below $[Z] < -2.25$ but the \model{S-} cases show more particles with metallicities in the range $-2.25 < [Z] < -0.75$, indicating even greater mixing than the \model{D-} cases. 

The MDFs in the hot CGM, the rightmost column of Fig. \ref{fig:mfm_iso_metal}, follow similar trends to the those in the warm CGM. At $t = 0.25$ Gyr for the \model{None} case, the peak due to enriched ($[Z] \approx -0.3$) ISM gas entering the halo is prominent.  There is also a small spread at $[Z] = -3.0$ for reasons already noted.

At $t \geqslant 0.5$ Gyr, the \model{None} case undergoes slight evolution and  by $t = 0.75$ Gyr, a mild positive slope develops in the range $-2.5 \leqslant [Z] \leqslant -0.75$.  The \model{D-} and \model{S-} cases follow near identical evolution at these later stages of the simulation, each with a spread in the distribution at $[Z] = -3.0$.  Here the differences between the \model{S-} and \model{D-} cases are minimal since the hot gaseous halo is turbulent and dominated by random motions, driven by the galactic outflows.    

Our isolated experiment demonstrates that by endowing particles with diffusivities based on the local fluid properties, we obtain significant differences in the ISM.  The constant-coefficient Smagorinsky model causes the MDFs in the ISM to rapidly tighten toward the mean value, whereas the dynamic model predicts diffusivities orders-of-magnitude smaller and the corresponding MDFs closely follow the \model{None} case.  Simultaneously, in the hot turbulent halo, the constant-coefficient and dynamic Smagorinsky models produce similar distributions by $t = 0.5$ Gyr due to the latter having higher values of \smagend.  Overall, for the constant-coefficient model, non-negligible shear in all gas phases causes the rapid diffusion of fluid properties, whereas the dynamic model allows different regions of phase space to undergo unique evolution in terms of the MDFs.  We stress the importance of the unique evolution of both phases of gas:  with a constant \smagend, it is impossible to capture the decreased mixing in the ISM while simultaneously capturing the high level of mixing in the hot turbulent halo.  The dynamic model provides an interesting avenue for follow-up study with zoom-in simulations, in order to gauge the impact on the ISM and CGM in a cosmological context.

\section{Cosmological volumes}
\label{sec:cosmo}

As we demonstrate in the previous section, turbulent mixing can alter the distribution of metals in various gas phases of an isolated disc experiment depending on the localisation of \smagend.  In a realistic cosmological environment, the evolution of galaxies is much more complex due to interactions between the galaxies and their environments.  These interactions include galaxy mergers, gas inflows and outflows, tidal interactions, ram pressure stripping, etc. -- all of which contribute to the production of turbulence in the galactic environments \citep{Iapichino2013, Schmidt2016}.  The resulting turbulence redistributes thermal energy, momentum, and metals, and must be included in numerical studies of galaxy evolution in order to have a self-consistent treatment of the physical models.

In contemporary Lagrangian-based numerical cosmological experiments, smoothed particle hydrodynamics (SPH) has been employed in simulation programs such as EAGLE \citep{Schaye2014}, OWLS \citep{Schaye2010}, and Romulus \citep{Tremmel2017}, whereas recently mesh-free finite mass (MFM) method has been employed in the MUFASA \citep{Dave2016c, Dave2017} simulations.  These experiments have produced a wealth of results for understanding galaxy evolution and gas properties (see \citealt{Somerville2015, Naab2016} for a summary), despite the fact that it is not possible for contemporary models to include all of the relevant physics\footnote{For several examples see \cite{Naab2016}.}.  Including sub-grid turbulent mixing could alter the results of such large-scale simulations.  Indeed, \cite{Tremmel2018} argue that turbulent mixing is critical for efficient redistribution of thermal energy released during active galactic nuclei episodes, and in previous sections, we showed that metal redistribution is also affected by turbulent mixing.  In models where the over-diffusive Smagorinsky model is used, the dynamic model can lead to differences in, for example, gas-phase metal abundances and hence, stellar abundances.  In principle, the differences in the manner and the rate at which metals are distributed could also affect the formation sites and population statistics of Population III stars and direct collapse seed supermassive black holes (see Section \ref{sec:introduction}).

In this section, we examine a set of cosmological simulations in order to test the effects of the dynamic model on the global gas enrichment levels and distributions.  In what follows, we adopt the MUFASA model \citep{Dave2016c} in combination with the diffusion models we describe in Table \ref{tbl:modeldescription}.  We choose to use the MUFASA model partly because it is the only cosmological model that has been implemented using the MFM method at the present.  

The initial conditions were created using a modified version of \pkg{GRAFIC-2}\footnote{\url{http://web.mit.edu/edbert/}} \citep{Bertschinger2011} and the parameters describing our simulations are listed in Table \ref{tbl:cosmoparams}.  In the following subsection, we briefly describe the MUFASA models, and point interested readers to \cite{Dave2016c} and \cite{Dave2017} for a more detailed explanation of the physical models. In Sections \ref{sec:cosmophasefracs} \& \ref{sec:cosmophasedists} we examine the global gas-phase metallicity fractions and the metal distribution functions (MDFs), respectively, from the simulation suite.

\begin{table}
	\caption{Parameters for the cosmological simulations.  We use the \protect\cite{Ade2016} cosmological model}.
	\label{tbl:cosmoparams}
 	\begin{tabular}{ll}
  		\hline
  		Simulation Parameters \\
  		\hline
  		L & 25 Mpc h$^{-1}$\\
  		N & 2 $\times$ 256$^3$ \\
  		m$_\mathrm{g}$ & 1.26 $\times$ 10$^7$ $\Msun$ h$^{-1}$\\
  		m$_\mathrm{dm}$ & 6.88 $\times$ 10$^7$ $\Msun$ h$^{-1}$\\
  		$\epsilon_\mathrm{soft,min}$ & 0.5 kpc h$^{-1}$\\
  		z$_\mathrm{init}$ & 70 \\
  		T$_\mathrm{init}$ & 59 K\\
  		\hline
  		Cosmological Model \\
  		\hline
  		$\Omega_\mathrm{m,0}$ & 0.308\\
  		$\Omega_\mathrm{\Lambda,0}$ & 0.692\\
  		$\Omega_\mathrm{b,0}$ & 0.048\\
  		$h$ & 0.678\\
  		$\sigma_\mathrm{8}$ & 0.815\\
  		$n_\mathrm{s}$ & 0.968\\
  		\hline
	\end{tabular}
\end{table}

\subsection{MUFASA}
\label{sec:mufasa}

The MUFASA simulations include the sub-grid models we described in Section \ref{sec:isogalaxymodels} and sub-sections therein, with some modifications to the star formation recipe and feedback described in the following sub-sections.

\subsubsection{Star formation}
\label{sec:mufasacoolsfr}

Star formation is based on the molecular gas model of \cite{Krumholz2009}, and the implementation dynamically calculates the fraction of molecular hydrogen in gas particles, $f_\mathrm{H_2}$, based on the gas surface density and the metallicity.  For more precise details, see \cite{Dave2016c} and \cite{Krumholz2009}.  The star formation rate follows,

\begin{equation}
\frac{\mathrm{d}\rho_*}{\mathrm{d}t} = \epsilon_* \frac{\rho f_\mathrm{H_2}}{t_\mathrm{dyn}},
\label{eq:mufasasfr}
\end{equation}

\noindent where $t_\mathrm{dyn} = (G\rho)^{-1/2}$ is the local dynamical time, $\rho$ is the density of the gas, and $\epsilon_* = 0.02$ is the efficiency of star formation \citep{Kennicutt1998}.  The critical density of star formation is taken at $n_\mathrm{*,crit} = 0.2$ cm$^{-3}$ following \cite{Dave2016c}. 

For our cosmological simulations, we do not employ the sub-grid interstellar medium (ISM) model of \cite{Springel2003a}, however we still require that the Jeans mass is resolved.  Therefore, in order to prevent numerical fragmentation at high densities, an artificial pressure is applied above a density

\begin{equation}
n_\mathrm{th} = \frac{3}{4\pi \mu m_\mathrm{p}} \bigg(\frac{5k_\mathrm{b} T_\mathrm{0}}{G\mu m_\mathrm{p}}\bigg)^3 \bigg(\frac{1}{N_\mathrm{ngb} m_\mathrm{g}}\bigg)^2,
\label{eq:mufasamaxdens}
\end{equation}

\noindent where $m_\mathrm{g}$ is the gas particle mass, $\mu = 1.22$, $T_\mathrm{0} = 10^4$ K, and $N_\mathrm{ngb} = 64$ is the number of neighbours.  The pressure is applied in the form of a minimum temperature \citep{Teyssier2011, Dave2016c},

\begin{equation}
T_\mathrm{JMT} = T_\mathrm{0} \bigg(\frac{n}{n_\mathrm{th}}\bigg)^{1/3}.
\label{eq:mufasajmt}
\end{equation}

\subsubsection{Feedback}
\label{sec:mufasafeedback}

In Section \ref{sec:isogalaxyfeedback}, we described the decoupled-wind model for massive star feedback in addition to feedback from supernova type-Ia (SNIa) and asymptotic giant branch (AGB) stars.  We use the same in the following experiments.

The simulations we present here do not include explicit active galactic nuclei (AGN) feedback.  AGN feedback is thought to be necessary to prevent excessive cooling and quench star formation in massive systems (see \citealp{King2015} for a recent review), but our smaller simulation volumes do not include many such systems.  We do, however, include an effective AGN feedback model from the MUFASA simulations that mimics the action of AGN feedback, and suppresses cooling of the diffuse halo gas.  Specifically, gas that is not self-shielded in halos with $M_\mathrm{halo} > M_\mathrm{q}$ (where $M_\mathrm{q} = (0.96 + 0.48 z)\times 10^{12}\; \mathrm{M}_\mathrm{\odot}$) is heated to 20\% above the virial temperature of the halo \citep{Mitra2015}.  The virial temperature follows \citep{Balogh1999, Voit2005},

\begin{equation}
\label{eq:virialtemp}
T_\mathrm{vir} = 9.52\times 10^7 \bigg(\frac{M_\mathrm{halo}}{10^{15} \mathrm{M}_\odot \mathrm{h}^{-1}}\bigg)^{2/3}\; \mathrm{K}.
\end{equation}

\subsection{Gas phases}
\label{sec:cosmophases}

\begin{table}
 \caption{We separate gas in our cosmological simulations into five phases: the interstellar medium (ISM), cool circumgalactic medium (CGM), hot halo gas (HHG), warm-hot intergalactic medium (WHIM), and cool diffuse gas (DIFF). $\rho_*$ is the star formation threshold, $n_{*,\mathrm{crit}} = 0.2$ cm$^{-3}$ and we give $\rho_\mathrm{bound}$ in equation~(\ref{eq:bounddensitycut}).}
 \label{tbl:phasedescription}
 \begin{tabular}{lll}
  \hline
  Name & Density Range & Temperature Range \\                            
  \hline
  ISM & $\rho > \rho_*$ & Any \\
  CGM & $\rho_* > \rho > \rho_\mathrm{bound}$ & Below equation~(\ref{eq:cgmhotline}) \\
  HHG & $\rho > \rho_\mathrm{bound}$ & Above equation~(\ref{eq:cgmhotline}) \\
  WHIM & $\rho < \rho_\mathrm{bound}$ & $T > 10^5$ K \\
  DIFF & $\rho < \rho_\mathrm{bound}$ & $T < 10^5$ K \\
  \hline
 \end{tabular}
\end{table}

We define five separate gas phases for the following subsections, and examine their properties in a global sense over the entire simulation volume.   We give a summary in Table~\ref{tbl:phasedescription}.  The definitions are largely from \cite{Dave2010}, except for the definition of the gas associated with the circumgalactic medium of galaxies.  For density, we cut the gas phase space using two thresholds $\rho_\mathrm{bound}$ and $\rho_*$, with $\rho_\mathrm{bound} = \rho_\mathrm{bound}(z)$ following,

\begin{equation}
\frac{\rho_\mathrm{bound}(z)}{\Omega_\mathrm{b}(z)\rho_\mathrm{c}(z)} = 6\pi^2 \bigg(1 + 0.4093 \bigg(\frac{1}{\Omega_\mathrm{m}(z)} - 1\bigg)^{0.9052}\bigg) - 1,
\label{eq:bounddensitycut}
\end{equation}

\noindent where

\begin{equation}
\Omega_\mathrm{m}(z) = \frac{\Omega_\mathrm{m,0} (1 + z)^3}{\Omega_\mathrm{m,0}(1 + z)^3 + \Omega_\mathrm{\Lambda,0}}\; ,
\label{eq:boundfactor}
\end{equation}

\begin{equation}
\Omega_\mathrm{b}(z) = \frac{\Omega_\mathrm{b,0} (1 + z)^3}{\Omega_\mathrm{m,0}(1 + z)^3 + \Omega_\mathrm{\Lambda,0}}\; ,
\label{eq:omegabaryonz}
\end{equation}

\noindent $\rho_{c}(z) = 3(H(z))^2 / (8\pi G)$, and $H(z) = H_\mathrm{0} \sqrt{\Omega_\mathrm{m,0}(1 + z)^3 + \Omega_\mathrm{\Lambda, 0}}$.  For the second density cut, we adopt $\rho_* = 4.4\times 10^{-25}$ g cm$^{-3}$, the star formation density threshold.  We have also applied a single temperature cut at $T_\mathrm{5} = 10^5$K to separate the warm-hot intergalactic medium (WHIM) from the cold diffuse gas (DIFF).  Following \cite{Torrey2017}, we apply a cut to distinguish the hot halo gas (HHG) from the cool circumgalactic medium (CGM),

\begin{equation}
\log{\bigg(\frac{T}{10^6\; \mathrm{K}}\bigg)} = 0.25 \log{\bigg(\frac{n}{405\; \mathrm{cm}^{-3}}\bigg)},
\label{eq:cgmhotline}
\end{equation}

\noindent where $T$ is the gas temperature, and $n$ is the gas density.  We define the HHG to be above the temperature threshold in equation~(\ref{eq:cgmhotline}), and the CGM to be below.

It is important to note that these density and temperature cuts do not distinguish gas domains that precisely correspond to their associated galactic or inter-galactic regions.  At high redshift ($z \gtrsim 5$), most of the gas classified as the WHIM phase is spatially located in the region that mostly corresponds to the CGM/HHG, and corresponds to outflowing galactic winds from early star formation, which gives rise to low-density, high-temperature gas.  Similarly, a fraction of the gas classified as the CGM at $z \gtrsim 1$ is in the cores of the cosmic filaments.  However, by $z = 0$, the majority of what we consider the CGM is indeed inside the halos,  and the cores of the filaments eventually end up in the WHIM phase.  This introduces a transition from CGM to WHIM gas that might not be obvious at first glance.  We have not tried to address these trends or optimise the phase cuts because the present study does not pertain to the evolution of the phases \textit{per se}, but rather the effect of differences in mixing strength on approximately physical phase-space cuts.

\subsection{Global gas evolution}
\label{sec:cosmophasefracs}

\begin{figure*}
	\centering
	\includegraphics[width=0.9\linewidth]{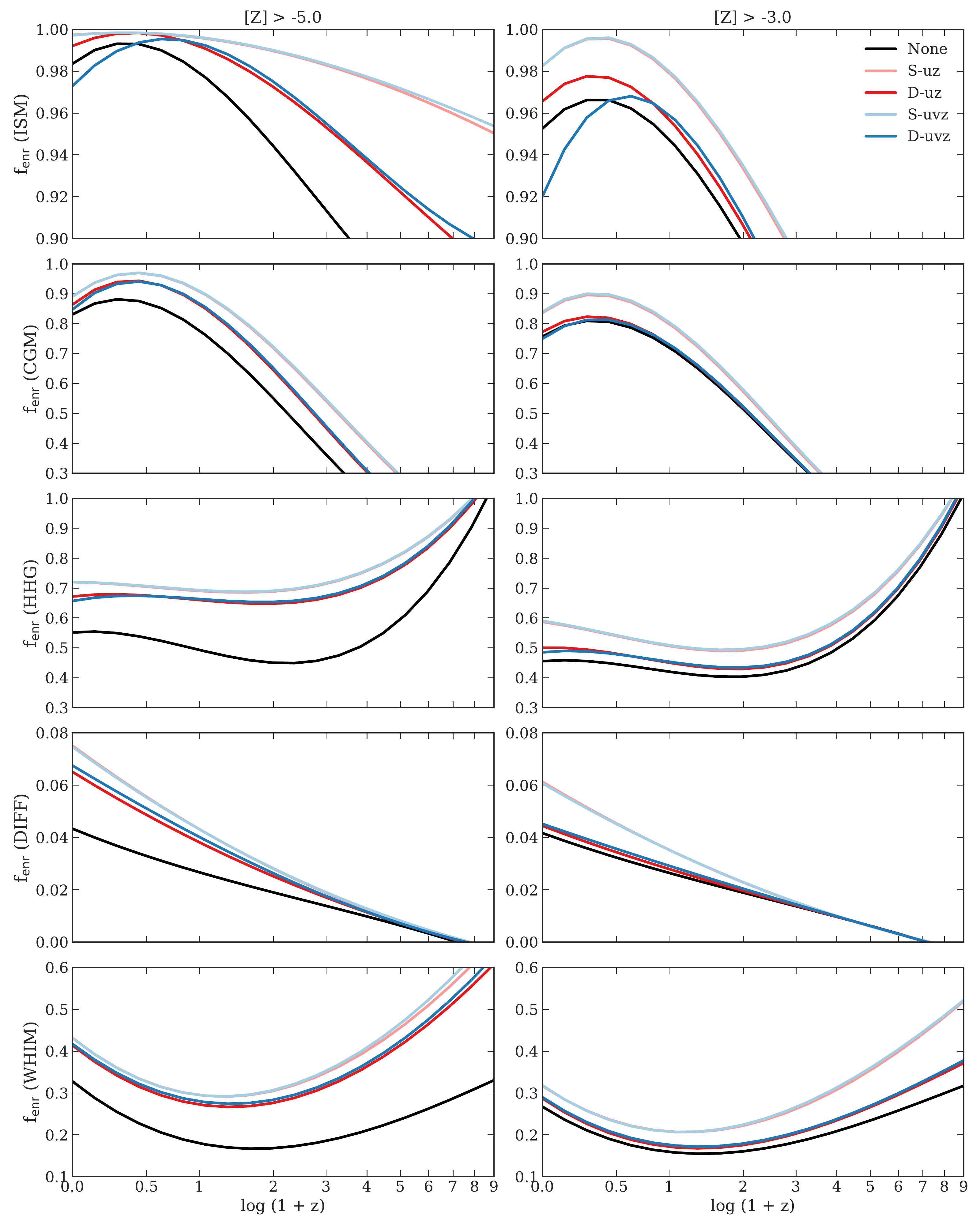}
	\caption{The enriched fraction as a function of redshift, defined as the mass fraction of gas with $Z > 10^{-5}$Z$_\mathrm{\odot}$ (left) and $Z > 10^{-3}$Z$_\mathrm{\odot}$ (right) in each phase.  The phases are, from top to bottom: interstellar medium (ISM), cool circumgalactic medium (CGM), hot halo gas (HHG), cold diffuse gas (DIFF), and the warm-hot intergalactic medium (WHIM).  Dynamic diffusion results in higher enriched fraction at the $[Z] > -5$ level, while maintaining a similar fraction to the no-mixing case above $[Z] > -3$.  The constant-coefficient Smagorinsky model increases the enriched fractions at all metallicity cuts.}
	\label{fig:mfm_global_fracs}
\end{figure*}

Fig. \ref{fig:mfm_global_fracs} shows the enriched fraction as a function of redshift, i.e. the ratio of the enriched gas mass to the total gas mass in each phase for two metallicity cuts: $[Z] > -5.0$ (left) and $[Z] > -3.0$ (right).  We use $[Z] \equiv \log{Z/Z_\mathrm{\odot}}$ as a proxy for metallicity, where $Z$ is the mass fraction of metals in a gas particle, and $Z_\mathrm{\odot} = 0.02$ is the solar metal mass fraction \citep{Anders1989}.  The rows represent the five phases defined in Table~\ref{tbl:phasedescription}.  

We first focus on the $[Z] > -5.0$ cut and start by examining the ISM results in the top row of Fig. \ref{fig:mfm_global_fracs}.  In the \model{None} case, the fraction of gas enriched to $[Z] > -5.0$ exceeds 90\% at $z = 4$, reaches a peak at $z = 0.5$, and then very slightly downturns by $z = 0$ due to accretion of low-metallicity gas.  Note that, in the following discussion, the differences between models are more important than the absolute values.  In the \model{S-} models, gas is enriched much earlier and we see 95\% of gas above $[Z] > -5.0$ by $z = 9$.   We do not observe the same slight downturn as in the \model{None} case.  This is not surprising.  The over-diffusive nature of the \model{S-} model leads to a reduced fraction of low-metallicity gas.  By $z = 0$, 100\% of the gas is above $[Z] > -5.0$ in the \model{S-} models.  Until $z \sim 1$, the \model{D-} models show enrichment levels intermediate between the \model{None} and \model{S-} cases.   At $z < 1$, the \model{D-uz} case continues this trend while the \model{D-uvz} model exhibits the biggest downturn. 

In the CGM, the second row in Fig. \ref{fig:mfm_global_fracs}, we notice similar trends to the ISM, at enrichment levels $[Z] > -5.0$.  For the \model{None} case, the gas is 30\% enriched at $z = 3$, 85\% (the peak) at $z = 0.5$, followed by a slight downturn.  The \model{S-} models follow the same qualitative trend but are at a constant 10\% above the \model{None} case, while the \model{D-} models remain in between the \model{S-} and \model{None} curves at all times. 

We do not discuss the details of the trends in the following three rows: the HHG, DIFF, and WHIM phases, respectively, but include them for completeness.  In these gas phases, the differences between mixing models are qualitatively the same as the CGM, whereas the \model{S-} and \model{D-} cases show significant increased enrichment above $[Z] > -5.0$.

Now we focus on the higher metallicity cut, $[Z] > -3.0$, in the ISM (top, right panel in Fig. \ref{fig:mfm_global_fracs}).  In the \model{None} case, gas is enriched over 90\% starting at redshift $z = 2$, reaches a peak at $z \approx 0.5$, and turns down by $z = 0$.  The trend for the $[Z] > -3.0$ gas is similar to that for the $[Z] > -5.0$ gas.  The normalisation of the curve is lower than at the $[Z] > -5.0$ threshold, as expected, since gas is enriched at higher metallicities later in cosmic evolution.  In the \model{S-} cases, gas is enriched above 90\% (for the $[Z] > -3.0$ cut) earlier compared to the \model{None} case, starting at $z = 3$, and remains above the \model{None} case at all times.   The \model{S-} cases show a similar downturn to the \model{None} case near $z \approx 0.5$.  Enrichment levels in the \model{D-uz} case remain slightly above the \model{None} case, but in \model{D-uvz}, there is a sharp downturn at $z \approx 0.75$ and the final enrichment level is below the \model{None} case.  The enrichment downturn in all of the mixing models is due a fresh supply of lower metallicity gas entering the medium, and the \model{D-uvz} model appears to amplify this effect.

We now discuss the enrichment levels above $[Z] > -3.0$ for the CGM in the right column of Fig. \ref{fig:mfm_global_fracs}, in the second row from the top.  In the \model{None} case, the gas is enriched above 30\% by $z \approx 3$, rises to a maximum of 80\% at $z \approx 0.4$, and slightly decreases to 75\% at $z = 0$.  Compared to the \model{None} case at the $[Z] > -5.0$ metallicity cut, the qualitative trend remains the same while the normalisation has decreased, for the same reason we describe above.  In the \model{S-} cases, the gas reaches 30\% enrichment levels somewhat earlier, by $z \approx 3.5$, and reaches a maximum enrichment level of 90\% at $z \approx 0.4$.  The maximum enrichment is 10\% higher in the \model{S-} cases compared to the \model{None} case showing that the constant-coefficient Smagorinsky model affects higher and lower metallicities equally.  The enrichment levels of the gas in the \model{D-} cases closely follows the \model{None} case for metallicities $[Z] > -3.0$, with only a slight divergence at $z = 0$. 

The HHG, DIFF, and WHIM phases follow the same qualitative trends as the CGM and, as noted above, we do not examine them in detail here.  However, these phases show that the dynamic model increases the gas enrichment levels above $[Z] > -3.0$ more so than in the CGM, but only slightly.  

As in the isolated galaxy experiment (see Section \ref{sec:isogalaxy}), the diffusivity plays a significant role in the enrichment levels of cosmic gas on a global scale.  The over-diffusive character of the constant-coefficient Smagorinsky model consistently results in the highest gas enrichment levels throughout the evolution of the simulations.  Interestingly, these same results also indicate that the dynamic model has the most effect on the metallicities in the range $-5.0 < [Z] < -3.0$ while leaving those above $[Z] > -3.0$ near the no-mixing level.   Early star formation and supermassive black hole formation are very sensitive to metallicities in this range (see Section \ref{sec:introduction}), and we have demonstrated that the dynamic model maximally affects those metallicities.  Although we do not resolve early star formation nor include supermassive black holes in our simulations, our results show that the dynamic model ought to be investigated further in cosmological simulations.

\subsection{Global gas-phase metallicity}
\label{sec:cosmophasedists}

\begin{figure*}
	\centering
	\includegraphics[width=\linewidth]{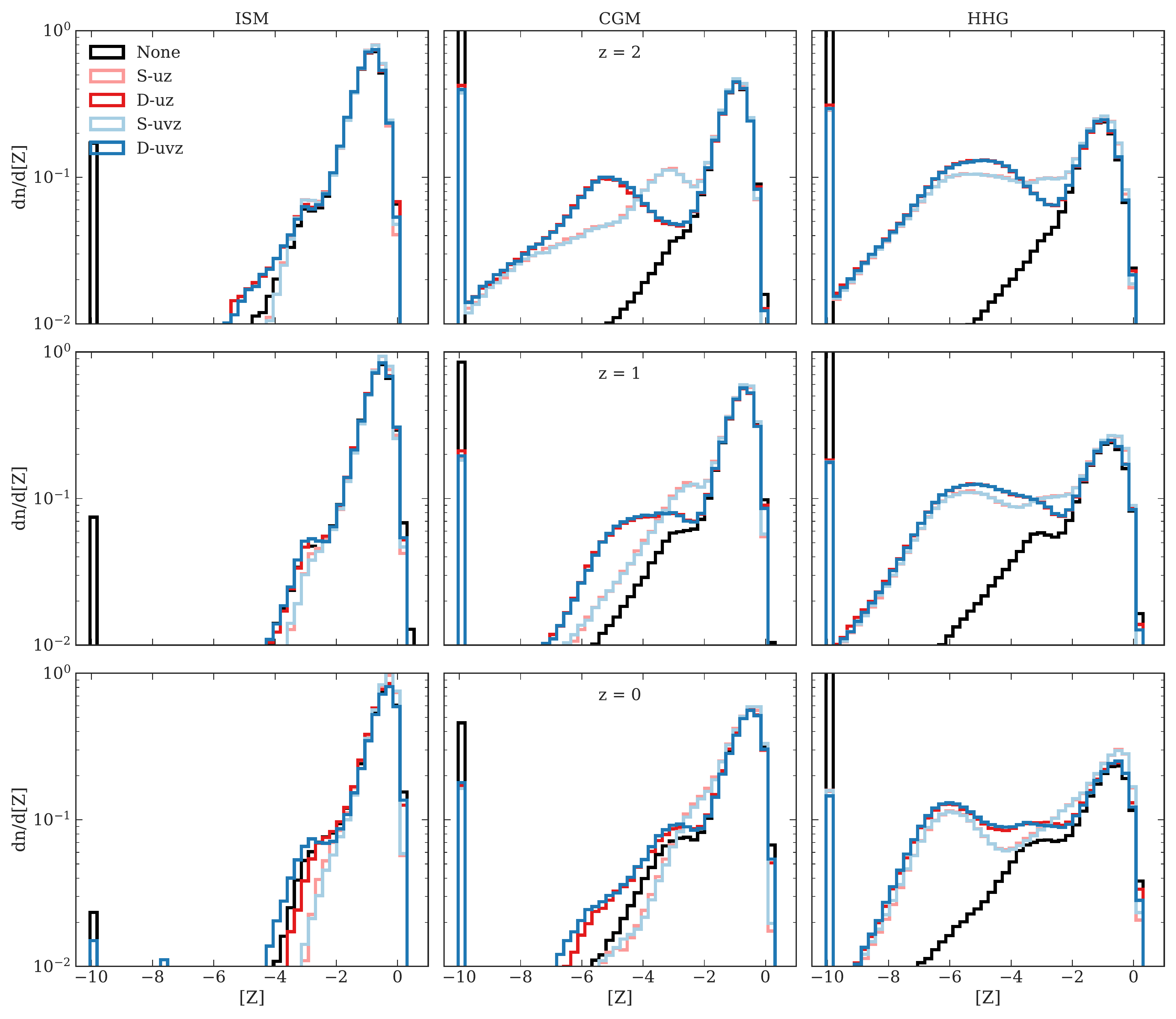}
	\caption{The metal distribution functions (MDFs) across three global gas phases (columns) in the simulation volume, at three separate evolutionary times (rows from top to bottom): $z = 2$, $1$, and $0$. We set a minimum metallicity of $[Z] = -10$ in our simulations, in order to show the abundance of unenriched gas.  The overall trends in each phase do not change significantly over time, yet comparing between mixing models reveal slight differences in the distributions of gas-phase metals.  Without metal mixing (\model{None}), galactic winds build MDFs in the non-star-forming gas that are similar to the distribution of the interstellar medium.  Turbulent mixing allows for further redistribution of metals once they arrive in the exterior media, altering the original distribution.  The diffusivity has a larger role in this method rather than the quantities diffused.}
	\label{fig:mfm_metal}
\end{figure*}

Metal distribution functions (MDFs) provide additional insight into the spatial redistribution of metals when compared with global enriched fractions, and we therefore investigate the MDFs of galactic gas at three separate evolutionary times: $z = 2, 1, 0$.  Our MDFs are probability density functions ($\mathrm{d}n/\mathrm{d}[Z]$, where $n$ is the number of particles) and are normalised such that the area under each curve is unity.  To facilitate plotting, we set a metallicity minimum of $[Z] = -10$ for all gas particles. Fig. \ref{fig:mfm_metal} shows the MDFs in the ISM, CGM, and HHG in the columns, left to right, and at $z = 2$, $1$, and $0$, from top to bottom.  The narrow, leftmost spike in the panels corresponds to the metallicity minimum. 

First, as we mention at the end of Section~\ref{sec:cosmophases}, it is important to note that the differences in the MDFs are not solely due to turbulent mixing.  There are transitions that occur across the sharp boundaries of the phase cuts we use.  In all of our mixing models, the cuts mostly affect the gas that falls within the CGM region of phase space and, in examining its spatial distribution at $z \gtrsim 1$, we find that the majority of the low metallicity ($[Z] < -3$) gas in the CGM is in the cosmic filaments.  This gas is not spatially associated with dark matter halos, but has the correct density and temperature to belong to the CGM phase.  Conversely, the gas in the ISM and HHG phase-space regions do correspond to what we consider their spatial counterparts, and the differences in their MDFs across the evolution of the simulation are dominated by turbulent mixing.

Turning to the distributions in Fig. \ref{fig:mfm_metal}, we first investigate the ISM.  The distributions across all models appear qualitatively similar, with slight differences at lower metallicities.  At $z = 2$, there is a peak at $[Z] \approx -1$ in all models, and in the \model{None} case, there is also near-pristine gas in the ISM.  The \model{D-} cases show a slightly extended tail covering the range $-6 \lesssim [Z] \lesssim -4$, compared to the \model{S-} and \model{None} cases, corresponding to the slight enrichment of the low-metallicity gas in the \model{None} case.  In the \model{S-} cases, the enrichment process is more efficient.  By $z = 1$, the tail tightens and all the distributions have negligible differences.  At $z = 0$, however, the \model{None} and \model{D-} cases share the same distribution whereas the \model{S-} models show a tighter distribution.  Additionally, there is much more gas at the minimum metallicity in the \model{None} case compared to both the \model{D-} and \model{S-} cases because metal enrichment of pristine particles in the \model{None} case only occurs when they are spatially adjacent to the stellar feedback sources.  In the \model{D-} and \model{S-} cases, any particle that is enriched acts as a local source of metals for neighbouring pristine particles, driving down the number of particles at the metallicity minimum. The trends here are similar to, and caused by, the same effect we see in the ISM of the isolated galaxy in Section \ref{sec:isogalaxy}, where the spread in the ISM MDF strongly depends on the diffusivity. 

Next, we examine the MDFs of the CGM.  In all models at $z = 2$, the gas in the CGM is a combination of the dense cores of the cosmic filaments and the cool, dense gas within dark matter halos.  In the \model{None} case, the MDF resembles the ISM distribution albeit with a more extended tail toward lower metallicities and more gas at the minimum metallicity.  Nonetheless, most of the enriched gas is above $[Z] \geqslant -6.0$.  The CGM distribution in the \model{None} case builds from the galactic winds transporting metals into the medium, sampling the underlying ISM distribution.  Additionally, SNIa and AGB stars contribute to varying metal distributions via the kernel-weighting procedure as gas spatially adjacent to the ISM is enriched at low levels. In the \model{None} case, the gas that is spatially in the cosmic filaments is at the metallicity minimum.  The $z = 2$ \model{D-} models show a bi-modality with peaks at $[Z] \approx -1.0$ and $[Z] \approx -5.5$, and in the \model{S-} models we also see a bi-modality but with peaks at $[Z] \approx -1.0$ and $[Z] \approx -3.0$.  The higher metallicity peaks in the \model{D-} and \model{S-} cases correspond to the peak metallicity in the ISM whereas the lower peak in each case is due to the gas that is classified as the CGM, yet is spatially in the cosmic filaments.  The spatial location of the gas does not change the effect of varying turbulent mixing strength; at redshift $z = 2$, the increased diffusivity in the \model{S-} models leads to a $2$ - $3$ order of magnitude shift compared to the \model{D-} models in the secondary low metallicity peak, indicating that the metals are much more dispersed in the \model{S-} cases.  Comparing to the \model{None} case, the question arises as to why the \model{D-} and \model{S-} cases have a broad lower metallicity peak.  At high redshift ($z \gtrsim 2$), in the \model{D-} and \model{S-} cases, the metal-enriched galactic winds escape the galactic halos and under the action of turbulent mixing, contaminate the gas in the filaments.  This is evidenced by the lower amplitude spike at $[Z] = -10$ in these models.  This does not occur in the \model{None} case, because the particles are unable to exchange metals directly.

By $z = 1$ the fraction of CGM gas in the filaments has dropped as the dense filament cores are heated and enter the WHIM.  However, there is still a small fraction of gas associated with the filaments.  The \model{None} case shows a qualitatively similar distribution compared to $z = 2$ because the gas leaving the cosmic filaments is at the metallicity minimum.  In the \model{D-} cases, the previous peak at $[Z] \approx -5.5$ becomes a broad shelf between $-5.0 < [Z] < -2.0$.  In the \model{S-} cases there is also a shelf of gas at $[Z] \approx -2.0$ but the main peak dominates.  The strong bi-modality from $z = 2$ has disappeared by $z = 1$ in the \model{D-} and \model{S-} models partly because the filamentary structure is increasingly classified as the WHIM and partly because the metallicity of the gas particles continues increasing due to mixing.

We turn now to the MDF of the CGM at $z = 0$.  At this redshift, most of the gas in the CGM region of phase space is associated with dark matter halos.  The \model{None} case has a similar distribution to $z = 1$ and $z = 2$.  The \model{D-} MDFs have narrowed further, with the low extended metallicity shelf at $z = 1$ transforming into a tail that extends from  $[Z] \approx -7.5$ to $[Z] \approx -3.0$, and a small shelf at $[Z] \approx -3.5$.  Above $[Z] > -2.0$, the \model{D-} MDFs coincide with the \model{None} case.  In the \model{S-} cases, the distribution also tightens and the tail is approximately a power law from $[Z] \approx -6.0$ to $[Z] \approx -1.0$, where the latter value is the peak of the distribution. The variations in the MDFs between the \model{None} case and both the \model{D-} and \model{S-} cases represent enriched fractions that have been altered via turbulent mixing rather than the aforementioned transitions between phases.

Shifting focus to the HHG at $z = 2$, we find that the gas in this phase is spatially associated with the dark matter halos.  We emphasise again that in the \model{None} case it is not possible for gas particles to exchange metals directly and consequently, the metal distributions only change via direct enrichment, or if the gas from the ISM reaches the phase under consideration via winds or via gas cooling from the intergalactic medium.  We see that the \model{None} case shows a similar distribution to both the ISM and the CGM, albeit with a slightly broader low-metallicity tail.  Like the $z = 2$ CGM distributions, the \model{D-} MDFs for the HHG show a bi-modality with peaks at $[Z] \approx -5.0$ and $[Z] \approx -1.0$ and a valley between $-4.0 < [Z] < -2.0$.  Unlike the CGM MDF, the HHG low-metallicity peak is broader and it is not due to gas transitioning from the HHG phase, but rather it is due to inflowing, low-metallicity gas mixing with the already enriched gas in the halos.  The \model{S-} cases share the peak at $[Z] \approx -1.0$, but the distribution is flat between $-6.0 < [Z] < -2.0$, before dropping-off toward low metallicities, in lockstep with the \model{D-} results.  The gap that is apparent in the \model{D-} cases has disappeared, indicating that in the \model{S-} cases there is much more gas with metallicities in the range $-4.0 < [Z] < -2.0$ than in the \model{D-} cases, which is not surprising given the over-diffusive nature of the constant-coefficient Smagorinsky model.

At $z = 1$ in the HHG, the \model{None} case remains unchanged except for a slight increase in the amount of gas near $[Z] \approx -3.0$, and the tail has extended slightly toward lower metallicity.  The latter is due to less enriched gas from the WHIM and DIFF phases accreting onto halos, diluting the metal distribution, and this dilution continues through to $z = 0$.  In the \model{D-} and \model{S-} cases, the low metallicity feature (peak/shelf and extended tail) has shifted slightly toward higher metallicity as turbulent mixing redistributes the metals from the highly enriched particles.

Next, we examine the HHG at $z = 0$.  The enrichment level, in the \model{None} case, is decreasing as evidenced by the tail of the MDF due to less enriched gas from the WHIM and DIFF phases accreting onto halos, diluting the metal distribution.  In the \model{D-} and \model{S-} cases, there are coincident peaks in the MDFs at $[Z] \approx -6.0$, although the \model{D-} cases show more gas at lower metallicities.  

Comparing across phases, the MDFs of the CGM gas are more sensitive to non-zero turbulent mixing strength than the ISM or the HHG.  For instance, even though the bi-modalities at $z = 2$ in the CGM MDFs, in the \model{S-} and \model{D-} cases, disappear by $z = 0$ due to the CGM to WHIM transition,  the dynamic model results in a residual extended tail in the CGM MDF as metals mix throughout the medium, whereas the constant-coefficient Smagorinsky model tightens the distribution in this phase and the gas metallicities rapidly approach the mean.  The peak in the CGM moves toward higher metallicities as gas flows between phases (CGM to WHIM).

The consequences of not including sub-grid turbulent mixing, regardless of the diffusivity, are clear -- complex structure in the MDFs is highly dependent on the mixing strength.  This is in contrast to \cite{Su2016} and \cite{Escala2017}, who found that turbulent metal mixing strength had low-level effects in their simulated ISM.  We posit that the low level effects were due the authours use of a constant value of \smagend, rather than localising the mixing strength to the appropriate regions.  We do not see the aforementioned trends in the \model{None} case because it is not possible for gas particles to exchange metals in our simulations.  Our stellar feedback model drives a decoupled wind from the ISM and that wind samples the MDF in the ISM, building up a similar metal distribution in the other phases that cannot change over time, unless material flows between the phases or due to delayed SNIa and AGB stars, via the kernel-weighted enrichment procedure.  

One important difference between the model we present and realistic environments is that our winds do not mix as they free-stream out of the galaxies.  While the coupling and mixing strength of galactic winds is uncertain, the wind could redistribute thermal energy, momentum, and metals internally as a cohesive unit, even if they do not couple strongly to the surrounding medium \citetext{Huang \& Katz, private communication}.  In our model, once the winds reach the criteria for recoupling in our simulations, they are free to mix their fluid properties and this, subsequently, allows for diverse MDFs in the gas phases exterior to the ISM.

We do not investigate the details of individual galaxies here but note that the differences outlined above will impact the gas-phase and stellar metallicities of those systems.  Therefore, including the dynamic model with a more accurate estimation of \smag is necessary, moving forward, in order to capture the physical redistribution of metals in galaxy evolution.

\section{Conclusions}
\label{sec:conclusions}

All hydrodynamical methods that are used to investigate galaxy evolution, whether Lagrangian or Eulerian, require additional sub-grid thermal energy and momentum diffusion terms in order to account for sub-grid turbulence.  In Lagrangian methods, such as mesh-free finite mass (MFM) and smoothed particle hydrodynamics (SPH), metal diffusion is also required due to the inability of fluid elements to exchange metals by construction.  Most implementations use the constant-coefficient Smagorinsky model -- one that has been shown to be over-diffusive in almost all cases, especially laminar shear flows.

We implemented and investigated the impact of the localised dynamic Smagorinsky model on global gas-phase properties in a series of numerical experiments using the \model{GIZMO} code.  In the dynamic case, the model coefficient depends on the local turbulent flow conditions, hence on the spatio-temporal coordinates.  This is in contrast to the constant-coefficient Smagorinsky model where diffusivities depend directly on the magnitude of the velocity shear in the fluid.  Compared to the constant-coefficient Smagorinsky model, the dynamic model has been shown to produce more accurate representations of fluid mechanical experiments \citep{Kleissl2006, Kirkpatrick2006, Khani2015, Benhamadouche2017, Lee2017, Kara2018, Taghinia2018}.  While we focused on cosmological experiments, the dynamic model has applications to any numerical experiment involving turbulent astrophysical flows, including stellar interiors, planetary formation, and star formation.  Moreover, the method we describe in this paper, following \cite{Germano1991} and \cite{Piomelli1994}, is general and not only limited to Lagrangian hydrodynamics, but also applicable to the Eulerian cases (see \citealt{Schmidt2015}).

For the MFM method, we showed that the dynamic model improves the density contrast in subsonic turbulence, allowing higher and lower density regions at fixed mass resolution.  In an idealised Keplerian disc, an example of a laminar shear flow where the Smagorinsky model is known to be over-diffusive from basic analytic arguments, the dynamic model produced near-zero values of turbulent diffusivity.   When we included thermal energy and momentum diffusion, the lower diffusivities prevented the rapid break-up of the disc due to excessive angular momentum transport.  We observed similar minimised diffusivities in a Kelvin-Helmholtz instability experiment, where the constant-coefficient Smagorinsky model smoothed, and rapidly diffused, our metal tracer, whereas the dynamic model captured the fine level of mixing at the interface of the two fluids.

We also investigated an isolated, Milky Way-like galaxy in order to test the dynamic model in a more complex, but still controlled, environment.  The dynamic model in combination with momentum diffusion improved the stability of the gaseous disc compared to the constant-coefficient Smagorinsky model, and affected the spatial metal distributions, as indicated by the metal distribution functions (MDFs), as shown in Fig.~\ref{fig:mfm_iso_metal}.  Rapid star formation early in the evolution of the disc leads to a higher diffusivity of thermal energy, momentum, and metals, and the subsequent exponential decay of star formation lowers the diffusivity in the dynamic model significantly.  This results in a broader MDF in the ISM in the dynamic case, pointing toward less mixing in the ISM.  When using the constant-coefficient Smagorinsky model, the diffusivity remained high throughout the evolution of the disc because of its strong dependence on the fluid velocity shear.  We found similar variations in the circumgalactic medium (CGM) for both the dynamic and constant-coefficient Smagorinsky models that we attribute to the turbulence generated from stellar feedback increasing the diffusivity in both cases.  

We also examined the global gas enrichment fractions in a set of cosmological simulations.  Global gas enrichment fractions are important for the formation of Population III stars and supermassive black holes because they are theorised to be sensitive to the metal content in the gas out of which they form (see Section \ref{sec:introduction} for more details).  We found that the dynamic model lowers overall enrichment compared to the standard Smagorinsky model, and that it maximally impacts metallicities in the range $-5.0 < [Z] < -3.0$.  This is precisely the metallicity regime that constrains the formation sites of supermassive black holes and Population III stars \citep{Volonteri2010, Sarmento2016}. Specifically, the dynamic model increases the amount of gas above $[Z] > -5.0$ while maintaining the same enriched fraction of gas above $[Z] > -3.0$, compared to the no-mixing case.  The standard Smagorinsky model increased the enriched fraction at all metal thresholds and in all gas phases.

In our cosmological simulations without turbulent mixing, we found that each gas phase external to the ISM has a qualitatively similar MDF to the ISM itself.  Turbulent mixing allows for regions to mix their metals, and introduces additional structure in MDFs of each phase.  We found that the diffusivity had a significant impact on the MDFs of the ISM and CGM -- the dynamic model shows broader MDFs in both phases at $z = 0$.  In these regions, we found a bi-modality in the CGM at $z = 2$ which disappeared by $z = 0$ in both cases, yet more lower metallicity gas remained in the dynamic case.  Our broad density and temperature phase-space criteria led to the bi-modality, as we found low-temperature and dense gas in the cores of the cosmic filaments at $z \sim 2$.  These spatial regions were of lower metallicity, and eventually return to the WHIM phase by $z = 0$. The peaks of the bi-modality, however, depend on the diffusivity: the dynamic model produced more metal poor (by several orders of magnitude) gas than the constant-coefficient Smagorinsky model.

Finally, we briefly touch on our conclusions for SPH.  Most authors apply the constant-coefficient Smagorinsky model to SPH \citep{Wadsley2008, Shen2010, Shen2013, Williamson2016a, Tremmel2017, Wadsley2017} and only include thermal energy and metal mixing.  In reality, there are additional turbulent transport terms that are unique to SPH (see \citealt{DiMascio2017} for an introduction and derivation) that must be included\footnote{It is possible to apply the dynamic model to the transport terms in \cite{DiMascio2017}, further improving upon their work.}.  Introducing momentum diffusion (e.g. via turbulent mixing as in \model{D-uvz} and \model{S-uvz}) in SPH is problematic because of unknown interactions with artificial viscosity.  Additionally, by construction, the smoothing kernel in SPH acts to produce coherent flows rather than fine structure observed in mesh-free or grid methods.  When we introduced momentum diffusion into SPH, the results from all of the experiments in this study were amplified when compared to the MFM method, but the qualitative trends remained.  Specifically, in our cosmological experiments, we found that momentum diffusion with the constant-coefficient model causes a delay in the formation of the interstellar medium (ISM) by $\sim 1$ Gyr, compared to the \model{None} case.  The dynamic model reduces the delay, but not to the no-mixing case.  We attribute this to momentum diffusion and dissipation causing the Jeans mass to increase resulting in the damping of mass fluctuations.

While we note that the dynamic turbulent mixing model introduced here is a step forward in understanding the redistribution of fluid properties in Lagrangian codes, there are caveats that must be explored.  The dynamic model predicts the correct behaviour in supersonic flows, but we did not include compressive mixing terms into our equations of motion.  It may be that compressive sub-grid mixing models further improve super-sonic turbulence in the MFM method.  Also, we justified using the Smagorinsky model by assuming that the local equilibrium condition holds, where the kinetic energy transfer rate down the turbulent cascade is equal on all scales.  While the assumption is approximately true on average in the regimes we investigated \citep{Schmidt2016}, a fully-consistent turbulence model involves tracking the sub-grid kinetic energy via an additional transport equation that includes all of the necessary, higher-order, sub-grid scale terms \citep{Schmidt2015}.  However, the dynamic model mitigates the issue by inherently calculating the deviations from local equilibrium.  Furthermore, the approximations for filtering the fluid fields require care and attention.  While \citeauthor{Monaghan2011}'s filtering approximation (equation \ref{eq:xsphsmoothing}) holds on the singly-filtered quantities for variations on scales larger than the resolution scale $h$, doubly-filtered quantities may be over- or under-smoothed.  A more robust, efficient, filtering procedure will need to be derived specifically for Lagrangian methods in the highly-compressible case, for filtering scales larger than $h$.

In summary, the dynamic Smagorinsky model localises the strength of turbulent mixing to only turbulent regions of the flow.  This provides a turbulent mixing model that does not rely on pre-calibrated parameters -- therefore simultaneously allowing near-zero diffusion in laminar shear flows and the expected diffusion in turbulent flows.  The physical experiments to which we subjected the model show that the dynamic model significantly alters the MDFs of the ISM and CGM in a global sense.  In future work we will examine the extent of small-scale differences associated with dynamic diffusion, and its impact on galaxy properties.

\section*{Acknowledgements}

This research was enabled in part by support provided by WestGrid and Compute/Calcul Canada. DR and AB acknowledge support from NSERC (Canada) through the Discovery Grant program. DR thanks the organisers of the \textit{Computing the Universe: At the Intersection of Computer Science and Cosmology} conference in Oaxaca, Mexico for an invited talk, and also James Wadsley and Andrey Kravtsov for their recommendations at the conference that led to the further refinement of this research project.  DR also thanks Valentin Perret for the \pkg{DICE} code, and Fabrice Durier, Ondrea Clarkson, Austin Davis, and Maan Hani for many useful discussions during the course of this research.  Support for PFH was provided by an Alfred P. Sloan Research Fellowship, NSF Collaborative Research Grant \#1715847 and CAREER grant \#1455342, and NASA grants NNX15AT06G, JPL 1589742, 17-ATP17-0214.  We would also especially like to thank our referee, Wolfram Schmidt, for his contributions in improving the final version of this study.




\bibliographystyle{mnras}
\bibliography{bibliography} 



\appendix


\bsp	
\label{lastpage}
\end{document}